\documentclass[fleqn,usenatbib]{mnras}
\usepackage{newtxtext,newtxmath}

\usepackage[T1]{fontenc}

\usepackage{graphicx}	
\usepackage{amsmath}	
\usepackage{comment}
\usepackage{todonotes}
\usepackage{adjustbox}
\usepackage{geometry,array,float,caption}
\usepackage{subcaption}
\usepackage{etoolbox}

\title[Learning to detect RFI without seeing it]{Learning to detect RFI in radio astronomy without seeing it }

\author[Michael Mesarcik et al.]{Michael  Mesarcik,$^{1}$\thanks{E-mail: m.mesarcik@uva.nl}
Albert-Jan Boonstra,$^{3}$
Elena Ranguelova, $^{2}$
Rob V. van Nieuwpoort$^{1,2}$
\\
$^{1}$Informatics Institute, Science Park 904, University of Amsterdam, The Netherlands\\
$^{2}$eScience Center, Science Park 140, 1098 XG Amsterdam, The Netherlands\\
$^{3}$ASTRON, the Netherlands Institute for Radio Astronomy, Oude Hoogeveensedijk 4, 7991 PD Dwingeloo, The Netherlands
}

\date{Accepted XXX. Received YYY; in original form ZZZ}

\pubyear{2022}

\begin{document}
\label{firstpage}
\pagerange{\pageref{firstpage}--\pageref{lastpage}}
\maketitle

\begin{abstract}
Radio Frequency Interference (RFI) corrupts astronomical measurements, thus affecting the performance of radio telescopes. To address this problem, supervised segmentation models have been proposed as candidate solutions to RFI detection. However, the unavailability of large labelled datasets, due to the prohibitive cost of annotating, makes these solutions unusable. To solve these shortcomings, we focus on the inverse problem; training models on only uncontaminated emissions thereby learning to discriminate RFI from all known astronomical signals and system noise. We use Nearest-Latent-Neighbours (NLN) - an algorithm that utilises both the reconstructions and latent distances to the nearest-neighbours in the latent space of generative autoencoding models for novelty detection. The uncontaminated regions are selected using \textit{weak-labels} in the form of RFI flags (generated by classical RFI flagging methods) available from most radio astronomical data archives at no additional cost. We evaluate performance on two independent datasets, one simulated from the HERA telescope and another consisting of real observations from LOFAR telescope. Additionally, we provide a small expert-labelled LOFAR dataset (i.e., strong labels) for evaluation of our and other methods. Performance is measured using AUROC, AUPRC and the maximum F1-score for a fixed threshold. For the simulated HERA dataset we outperform the current state-of-the-art across all metrics. For the LOFAR dataset our algorithm offers both a 4\% increase in AUROC and AUPRC at the cost of increasing the false negative rate, but without any manual labelling.

\end{abstract}


\section{Introduction}
Radio Frequency Interference (RFI) is a growing concern for radio astronomy due to the proliferation of electronic equipment that depends on electromagnetic emissions. Radio frequency radiation from astronomical sources is extremely faint relative to emissions from man-made systems such as radars, telecommunication devices, large satellite constellations~\cite{Hainaut2020} and more. Despite international regulation to ensure radio-quiet zones and limit transmission power of emitters, there are still concerns about RFI hampering radio astronomy.

For this reason, approaches for RFI detection and mitigation have become a necessity in modern radio observatories. RFI pipelines are commonly deployed at telescopes performing RFI detection and mitigation in a post correlation setting. Traditionally, algorithms such as \texttt{CUMSUM}~\cite{Baan2004}, Singular Value Decomposition (SVD)~\cite{Offringa2010a}, Wavelet-based methods~\cite{Maslakovic} and \texttt{AOFlagger}~\cite{Offringa2012a} have been used. These RFI detection algorithms are widely implemented for real-time RFI detection at observatories around the world~\cite{Foley2016, VanHaarlem2013, Kildal1991,Sokolowski2015}. In effect all archived data from these instruments contain automatically generated RFI-masks which can be accessed with no additional cost. 

Recent advances in machine learning have made data-driven approaches unprecedentedly suitable for RFI-detection.  Most machine learning approaches to RFI detection have been based on supervised learning using U-net~\cite{Ronneberger2015} and its derivatives~\cite{Akeret2017,Sadr2020, Kerrigan2019,Yang2020}.  Research has shown that these are promising methods, significantly outperforming classical approaches. However, in reality supervised methods require significant amounts of expertly labelled time/frequency data, that is not available in practice due to the related cost.  

As a result, recent models are trained and evaluated using simulators or flags generated by classical methods, with limited experimentation on real expertly labelled datasets. This is problematic as the effectiveness of these methods on unseen data is difficult to measure and predict. Furthermore, recent machine learning-based methods have not been well integrated into telescope-pipelines as the cost of labelling is prohibitive to many instrument operators. 

To solve these problems we propose an unsupervised learning method based on the Nearest Latent Neighbours (NLN) algorithm~\cite{Mesarcik2021}. This approach leverages novelty detection to perform RFI detection. This is achieved using a generative model trained on uncontaminated (RFI free) data to detect \textit{novel} RFI contaminated emissions. Interestingly, this formulation is effectively the inverse of how existing deep learning-based methods are trained.

In this work, RFI is detected by measuring the difference between small sub-regions (patches) of spectrograms that are known to {\em not} contain RFI and the patches being evaluated. To select RFI-free data we break each spectrogram into a number of equally sized patches and use their associated AOFlagger-based flags to locate the instances which contain RFI. In doing so we do not have to incur the cost of extensive labelling as the AOFlagger masks are readily available. It must be noted that NLN can tolerate high false positive rates in the training masks due to the use of the inverse problem. However, under-flagging can cause undesired effects.  

We show that our method outperforms existing supervised models on several benchmarks, using less data for training. Furthermore, we demonstrate that if supervised state-of-the-art methods are trained with weak-labels they typically over-fit to the training data and do not generalise to unseen examples. Our approach does not suffer from these problems.

Additionally, as the landscape of RFI emissions changes over time we expect supervised methods to be continually retrained as future emitters occupy newer frequency bands. Conversely, the emissions from celestial bodies will remain fairly consistent over the same time-scale, effectively meaning that that our method will not have to undergo retraining.

We make the following contributions in this work: (1) a novel unsupervised learning-based approach to RFI detection in radio astronomy; (2) an evaluation of the effectiveness of using AOFlagger generated ground-truth for training of machine learning-based RFI detection algorithms, and (3) an expert labelled dataset that can be used for comparison and development of novel RFI detection algorithms. 

This paper begins with the critical discussion of the existing literature concerning RFI detection in radio astronomy in Section~\ref{sec:Related_Work}. Section~\ref{sec:Method} explains how the NLN algorithm was adapted to work for RFI detection. In Section~\ref{sec:Data} we explain our data selection strategy and outline the expert-labelled dataset used for evaluation of this work.  Finally, we present our results and conclusions in Section~\ref{sec:Results} and Section~\ref{sec:Conclusions}.

\section{Related Work}
\label{sec:Related_Work}
Machine learning for RFI detection is an actively researched field. Numerous works offer a variety of radio astronomy-specific modifications to improve accuracy of detection. Furthermore, machine learning-based anomaly and novelty detection have been applied extensively in astronomy for purposes of novel galaxy morphology detection, transient detection, exo-planet discovery and more. However, there have been few attempts to apply novelty detection to the RFI detection problem. In this section we document the latest astronomical developments in both RFI detection and novelty detection. 

In this work, we omit analysis of and comparison with classical techniques that rely on spatial filtering, high order statistics and subspace decomposition~\cite{Baan2001}. This is because these techniques typically require significant fine tuning on specific telescopes thus making comparison challenging. 

\subsection{RFI detection in the deep learning era}
Semantic segmentation is at the heart of the deep learning-based RFI detection, with U-Net~\cite{Ronneberger2015} and derivatives acting as the architectural backbone of recent research. The purpose of semantic segmentation is to determine the pixel-precise regions where a specific class exists -- in this case RFI. Architecturally, U-Net is a Convolutional Neural Network (CNN), with an encoder-decoder pair that share activations between the two stages. It is trained in a supervised manner, requiring pixel-level Boolean masks per spectrogram.

The first application of U-Net to radio astronomy-based RFI detection is reported in seminal work by~\cite{Akeret2017}. The network is trained and evaluated on the magnitude of spectrograms obtained from both simulated data and real data from a signal antenna from the Bleien Observatory~\cite{Chang2017}. 
Interestingly, the models are trained using masks obtained from a classical flagging approach. We show in Section~\ref{sec:Results} that this is not ideal, as supervised methods tend to over-fit to the weak-label based ground-truth. Additionally, this work makes use of the HIDE \& SEEK radio astronomical data simulator ~\cite{Akeret2017b}. We find the use of this simulator problematic because the ground truth needs to be determined by user-defined thresholds of the residual RFI maps as described by \cite{Sadr2020}. Due to this, other works such as \cite{Yang2020} do not describe the threshold used for evaluation making comparison extremely difficult. 

To counter-act the issue of over-fitting to the potentially incorrect labels we focus on the inverse problem. We train a model to represent all non-RFI signals, such that any deviation from the learnt representations is flagged as RFI. It must be noted that this approach depends on the assumption that the number of false positives of classical RFI detectors is higher than the number of false negatives. In other words, most RFI is flagged as RFI, but other features may be incorrectly flagged. A detailed analysis of this effect can be found in Section~\ref{sec:HERA_results}.  

To address the limitations of the HIDE simulator, we use the Hydrogen Epoch of Reionization Array (HERA) simulator~\cite{DeBoer2017}. It gives more a granular control over the simulated interference, enabling multiple classes to be generated with precise ground truth without specifying a threshold. Furthermore, we train our model on multiple observations obtained from the Low-Frequency Array (LOFAR) Long Term Archive (LTA)~\cite{VanHaarlem2013} and evaluate it on a subset of expert-labelled examples.   

\cite{Kerrigan2019} document the use and evaluation of a U-Net derivative for RFI detection on both real and simulated data from the HERA telescope~\cite{DeBoer2017}. The authors propose to include both the magnitude and phase as separate components of the model, for better  generalisation to alternative representations. It must be noted that both~\cite{Mesarcik2020b, Sadr2020} show that minimal improvements are obtained through using both the magnitude and phase representations of the complex visibilities. For this reason, we only use the magnitude-portion of our data for RFI detection. Additionally, we extend the evaluation done by~\cite{Kerrigan2019} by considering how supervised RFI detection algorithms generalise to unseen classes of RFI, what we refer to as Out-of-Distribution (OOD) RFI. 

Transfer learning~\cite{Tang2019} or domain adaptation~\cite{Farahani2021} have been shown to be powerful tools when there is limited labelled data. \cite{Sadr2020} show that it is possible to train a U-Net inspired architecture, called \textit{R-Net}, on simulated data and then using a small sample of expert labelled data adapt the model's domain from simulated to real world data. However, we find that R-Net does not offer significant improvements over the standard U-Net, when evaluating on real data without transfer learning, as shown in Section~\ref{sec:Results}. Similarly \cite{Yang2020} propose \textit{RFI-Net}, a residual modification of U-Net that offers better training stability when using a significantly deeper network. 

Generative models have also been used for RFI detection in the context of radio astronomy. In work by~\cite{Vinsen2019} it has been shown that Generative Adversarial Networks (GANs) can be used for RFI detection. However this research is limited in its evaluation and does not offer a practical way to obtain pixel-precise predictions of RFI. Finally, \cite{Vos2019} offer a significantly different paradigm for RFI detection using GANs. Here, the authors propose a source-separation approach that uses 2 separate generators to distinguish astronomical signals from RFI. However, this method requires significant supervision, as the model needs access to the mixture as well as the separated RFI and astronomical sources during training. We find this requirement prohibitive, as to obtain these source separations for real data is extremely costly. We show that generative models can be used without the cost of supervision, by treating novelty detection as a downstream task. 

\subsection{Novelty detection in radio astronomy}
Machine learning-based novelty and anomaly detection have gained significant attention in radio astronomy for a number of different use-cases. Topics such as transient detection~\cite{Malanchev2021}, the Search for Extra-Terrestrial Intelligence (SETI)~\cite{Zhang2019} and detecting outliers in radio galaxy morphology~\cite{Margalef-Bentabol2020} have been researched. 

Novelty detection in machine learning is typically done in a two-stage process, where a model is first fit to the in-lying classes and then a decision boundary is found in the learnt-representation space of the model. In work by \cite{Villar2021} outlying extra-galactic transients are detected using an Isolation Forest (IF)~\cite{TonyLiu2008}, whereas \cite{Lochner2020} propose using Local Outlier Factor (LOF)~\cite{Mudinas2020} for detecting anomalies in light-curves or dynamic spectra. To the best of our knowledge there have been no attempts at applying these techniques to RFI detection in the context of radio astronomy.  

The main problem with directly applying latent-variable novelty detection models to RFI detection is the resolution of the predictions. The models used for novelty detection in radio astronomy produce a scalar output for a single data point (spectrogram, radio image, etc). This is problematic as RFI detection algorithms must produce a Boolean mask of the pixel-precise regions of where RFI is located in a given input.  Therefore, we propose to use the Nearest-Latent-Neighbours (NLN) algorithm, an approach that combines both latent measures of difference as well as the pixel-precise reconstruction errors from a generative autoencoding model~\cite{Mesarcik2021}. This is in contrast to~\cite{Robishawa}, where particular features are extracted from each spectrogram to determine the novel differences in transmission power and frequency range. However, the method does not produce pixel-precise segmentation maps of the detected RFI. 

\vspace{-0.45cm}
\section{Method}
\label{sec:Method}
 \begin{figure*}
     \centering
     \includegraphics[width=\linewidth]{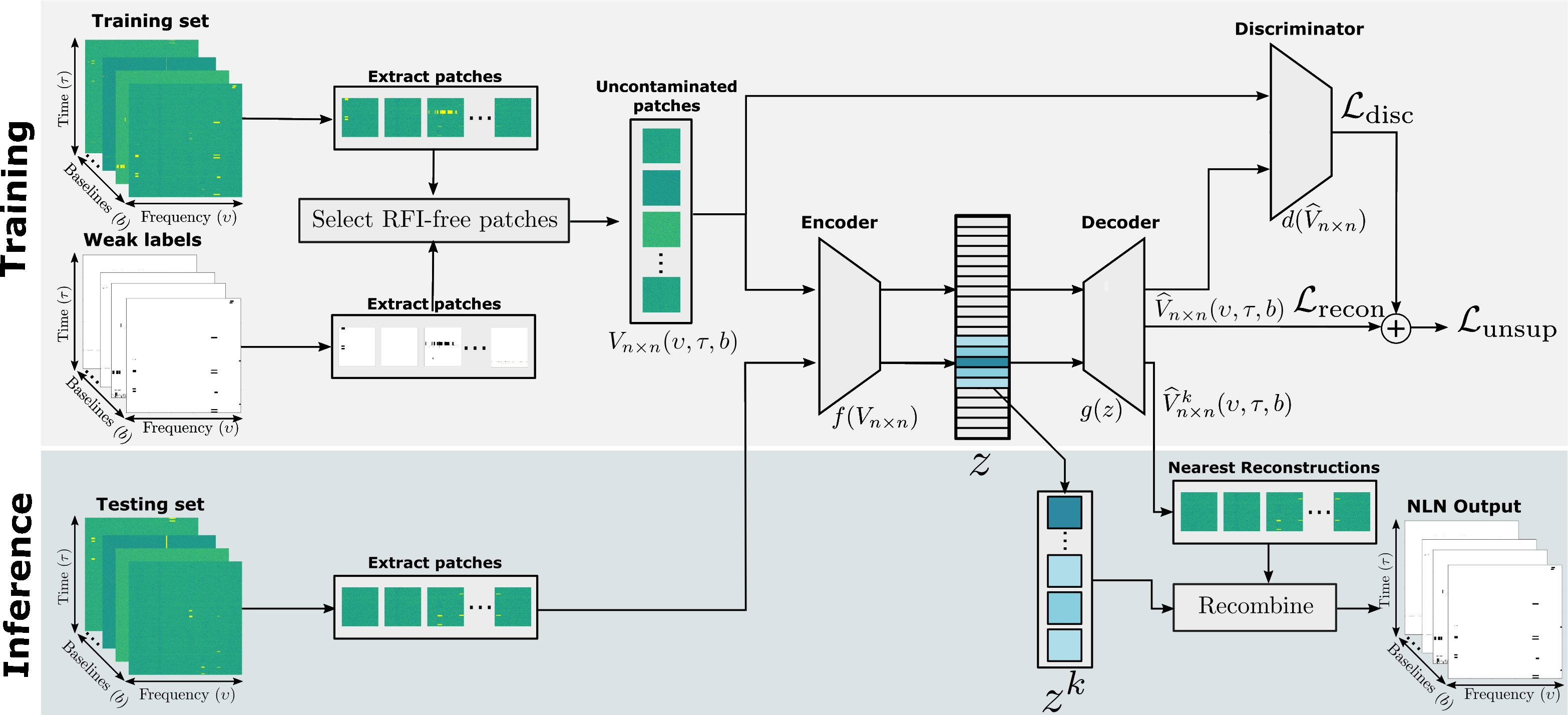}
     \caption{Block diagram of the training and inference procedures of NLN-based RFI detector. Here, we use a discriminative autoencoder as the backbone of the architecture. The top half of the figure shows the training procedure whereas the bottom illustrates how NLN is used for inference. The recombination of latent distances $z^k$ and nearest-reconstructions $\hat{V}^k(\upsilon, \tau, b$) is performed according to Equation~\ref{eq:NLN}}. 
     \label{fig:diagram}
 \end{figure*}
In this work we use NLN for RFI detection. This is motivated by several factors: (1) obtaining sufficient labels for supervised segmentation of RFI has a significant overhead; (2) existing supervised techniques over-fit to flags from classical methods such as AOFlagger, leading to sub-optimal performance on unseen data; (3) the ever-changing landscape of RFI requires continual labelling and training efforts to enable supervised approaches to capture new temporal and spectral RFI structures. 

We show that if the traditional RFI detection problem is inverted, we can effectively address the supervised RFI detection issues. All models and code used in this work are publicly available\footnote{\href{https://github.com/mesarcik/RFI-NLN}{https://github.com/mesarcik/RFI-NLN}}. 

\subsection{Model definition and training}
For some complex visibility $V(\upsilon, \tau,b)$ and the corresponding ground truth mask for the interference $G(\upsilon, \tau,b)$, the training objective for supervised RFI detection can be formulated as follows 
\begin{equation}
    \mathcal{L}_{\text{sup}} =  \min_{\theta_m} \mathcal{H}(m_{\theta_m}(V(\upsilon, \tau,b)), \; G(\upsilon, \tau,b)).
\end{equation}
Here $m$ is a function with learnable parameters $\theta_m$ and $\mathcal{H}$ is the entropy-based similarity between the model prediction and the ground truth. This problem is well-posed and has been used across multiple domains to effectively train classifiers. However it relies on learning a model of the RFI using ground truth labels, which are in practice hard to obtain. 

In this work, we train a model of everything other than the interference, so that we can perform RFI detection as a downstream task. This is done by first training a discriminative autoencoder, on $n\times n$-sized uncontaminated regions (also known as patches) of the visibility, $V_{n\times n}(\upsilon, \tau, b)$. We select these regions using the \textit{weak-labels} generated by a classical method such as AOFlagger.

First we define the encoder $f$ that maps from the visibility space $\mathbb{R}^2$ to a latent space $\mathbb{L}$, such that  

\begin{equation}
    \mathbf{z} = f_{\theta_f}(V_{n\times n}(\upsilon, \tau,b)),\quad f: \mathbb{R}^2 \rightarrow \mathbb{L}.
\end{equation}
Here $\mathbf{z}$ is a low-dimensional projection of the an $n\times n$ patch that contains no interference and $\theta_f$ are the learnable parameters of the encoder. Furthermore, we define the decoder $g$ that maps back from the low-dimensional projection to the visibility space, such that 

\begin{equation}
    \widehat{V}_{n \times n}(\upsilon, \tau, b) = g_{\theta_g}(\mathbf{z}), \quad g: \mathbb{L} \rightarrow \mathbb{R}^2
\end{equation}
where $\theta_g$ are the parameters of the decoder. We simultaneously train the encoder and decoder using the reconstruction loss,
\begin{equation}
    \mathcal{L}_\text{recon} = \min_{\theta_{f}, \theta_{g}}  \mathcal{H}(V_{n\times n}(\upsilon, \tau,b), \; \widehat{V}_{n \times n}(\upsilon, \tau,b)).
\end{equation}
We use Mean-Square-Error (MSE) to train a standard autoencoder. Typically, MSE-based reconstruction losses produce blurry outputs, which may affect the quality of the predicted RFI masks. The blurriness is a result of back-propagating the gradient from the average pixel-wise error, which prohibits the autoencoder from producing high-frequency details for given inputs. In order to counteract this problem, we define a discriminator, $d$, that acts as a regulariser on the decoder's output. This discriminative loss is back-propagated through the decoder, such that it learns to produce not only low frequency details but also the discriminative features~\cite{Srivastava2017}. Furthermore, the discriminator enables the autoencoder to be used as a generative model~\cite{Larsen2016}. The discriminator maps from $\mathbb{R}^2$ to a classification on the interval $[0,1]$. Effectively trying to determine if the input is generated or sampled from the original dataset. In this case, the original dataset is the uncontaminated patches selected using the weak-labels. The discriminative loss is given by 

\begin{equation}
    \begin{split}
         \mathcal{L}_\text{disc} = \min_{\theta_{d}} \; \mathbb{E} & [ \log ( d_{\theta_d}((V_{n\times n}(\upsilon, \tau,b)))]  + \\
        \mathbb{E} &[ \log(1-d_{\theta_d}( \widehat{V}_{n \times n}(\upsilon, \tau,b)  )) ].
    \end{split}
\end{equation}
We train the discriminator simultaneously with the decoder, such that the total loss is $\mathcal{L}_{\text{unsup}} = \alpha\mathcal{L}_\text{recon} + (1-\alpha)\mathcal{L}_\text{disc}$, where $\alpha$ is a hyper-parameter between 0 and 1 that  determines how much of an effect the discriminative and reconstruction losses have on the training respectively. An illustration of the training procedure for the discriminative autoencoder is shown in the top half of Figure~\ref{fig:diagram}.

\subsection{Nearest-Latent-Neighbours for RFI detection}
\begin{figure*}
     \centering
     \begin{subfigure}[b]{0.32\linewidth}
         \centering
         \includegraphics[width=\linewidth]{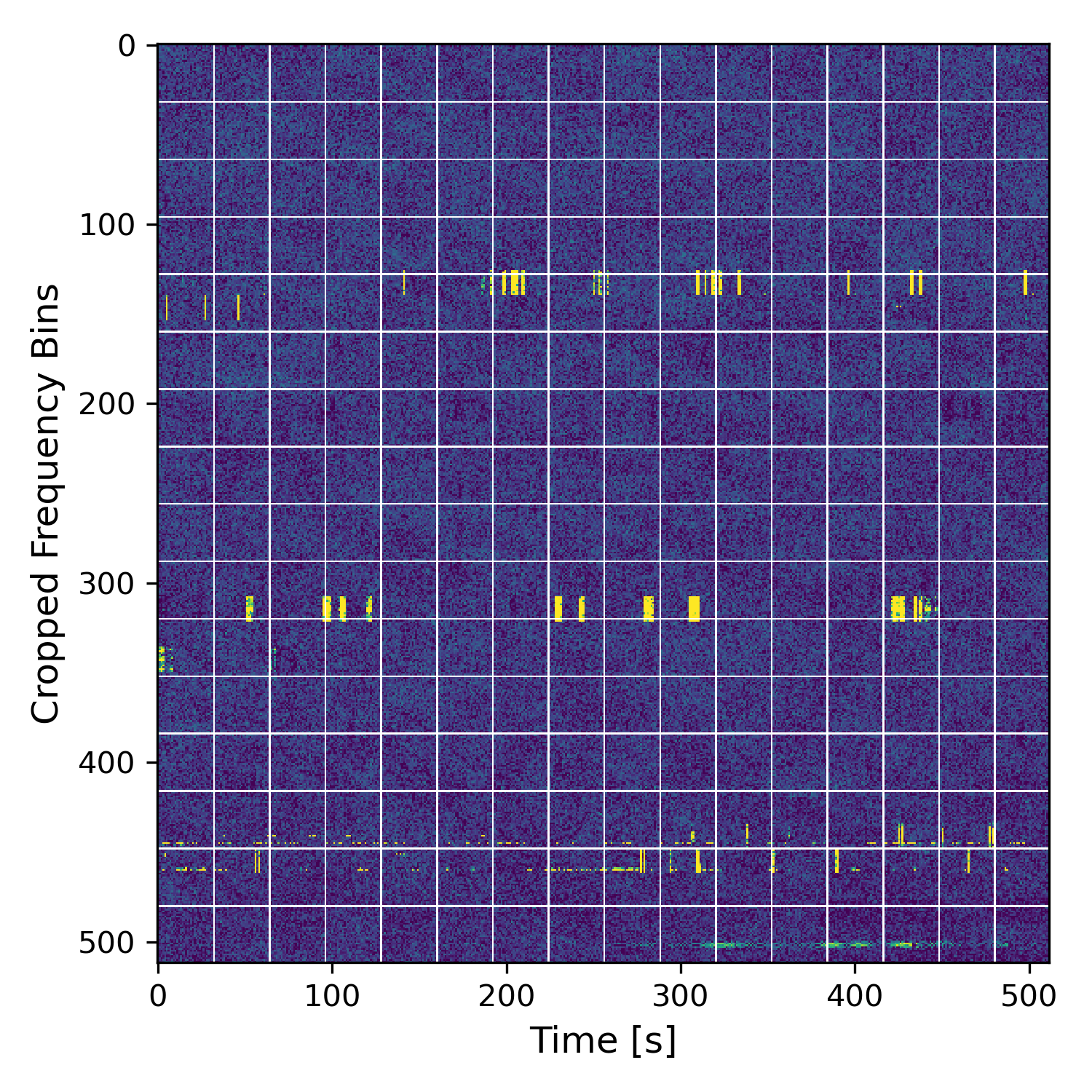}
         \caption{Input: $V(\upsilon, \tau,b)$}
         \label{fig:input}
     \end{subfigure}
     ~
     \begin{subfigure}[b]{0.32\linewidth}
         \centering
         \includegraphics[clip,width=\linewidth]{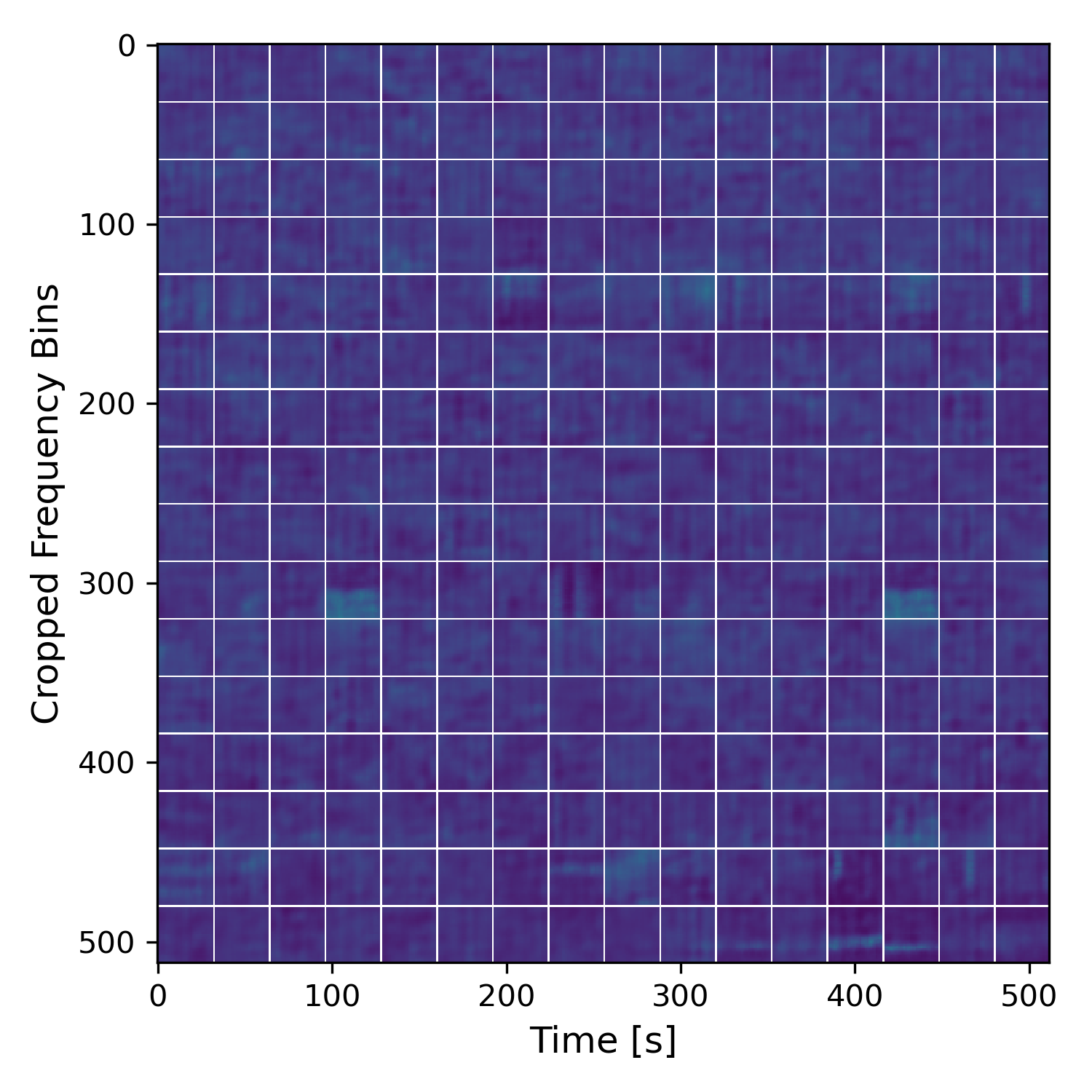}
         \caption{1$^{\text{th}}$ Nearest-Neighbour: $\hat{V}^1(\upsilon, \tau,b)$}
         \label{fig:neighbour_0}
     \end{subfigure}
     ~
      \begin{subfigure}[b]{0.32\linewidth}
         \centering
         \includegraphics[clip,width=\linewidth]{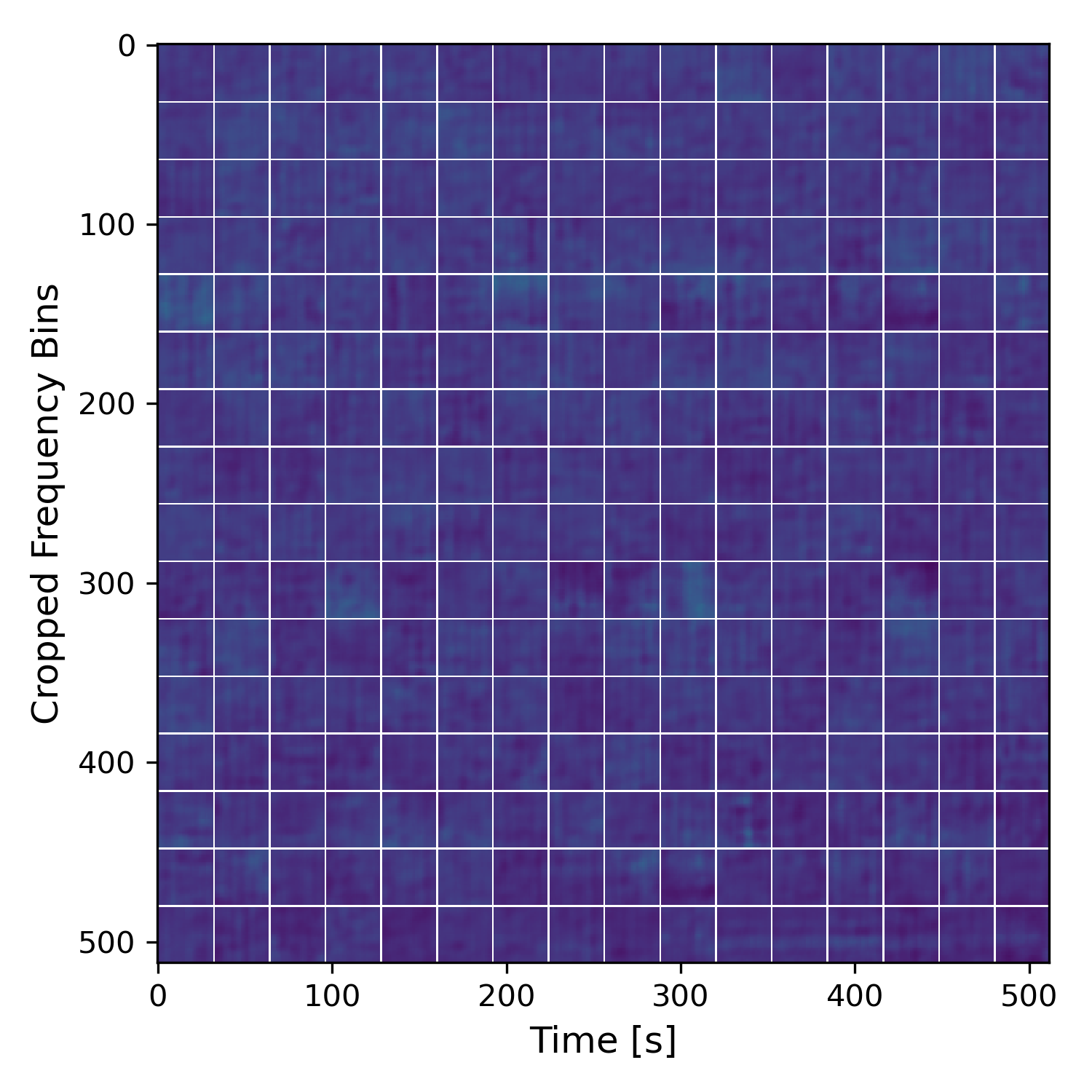}
         \caption{10$^{\text{th}}$ Nearest-Neighbour: $\hat{V}^{10}(\upsilon, \tau,b)$}
         \label{fig:neighbour_10}
     \end{subfigure}
     ~
     \begin{subfigure}[b]{0.32\linewidth}
         \centering
         \includegraphics[width=\linewidth]{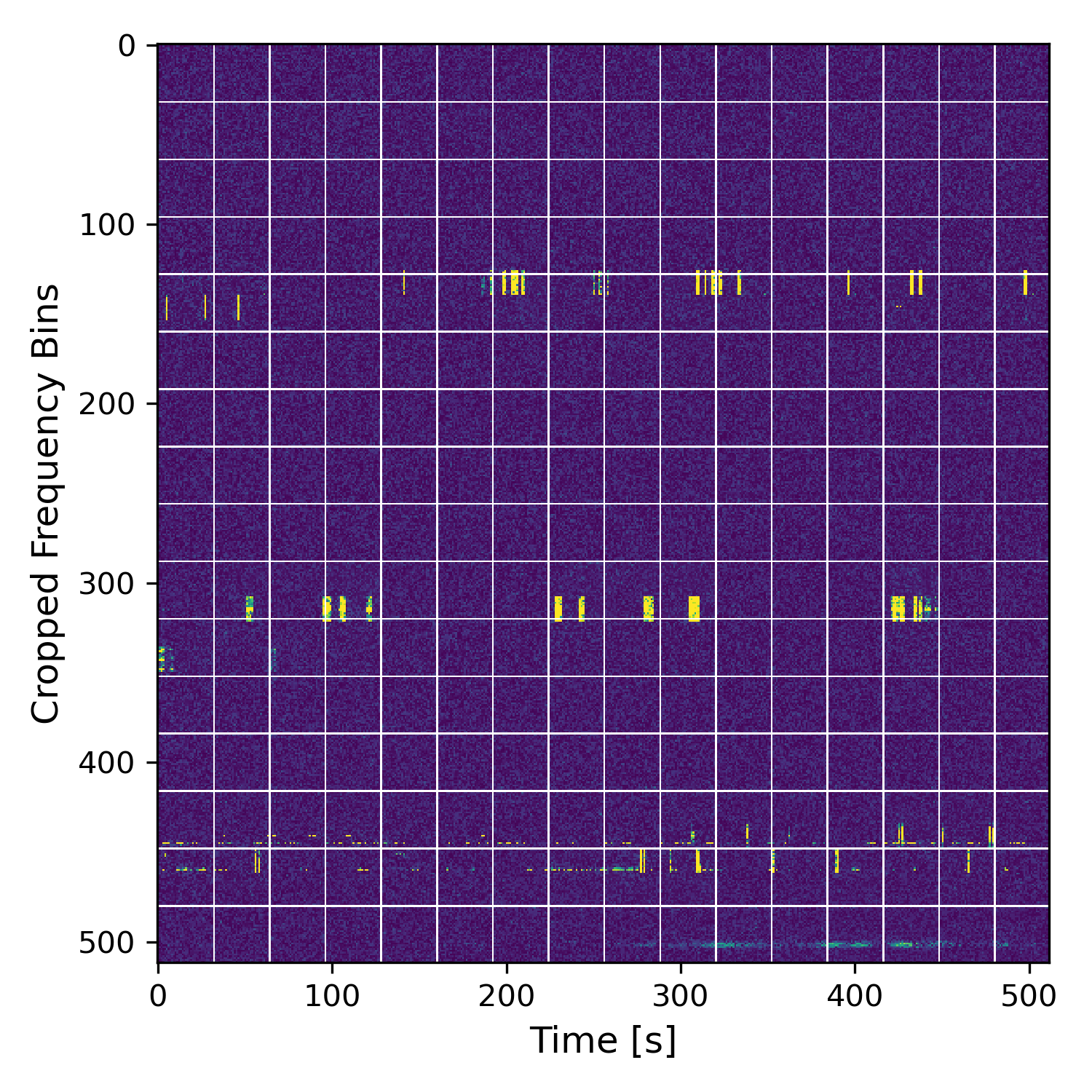}
         \caption{Error: $\dfrac{1}{K}\sum^K |V(\upsilon, \tau,b) - \hat{V}^k(\upsilon, \tau,b)|$}
         \label{fig:nln_error}
     \end{subfigure}
     ~
     \begin{subfigure}[b]{0.32\linewidth}
         \centering
         \includegraphics[width=\linewidth]{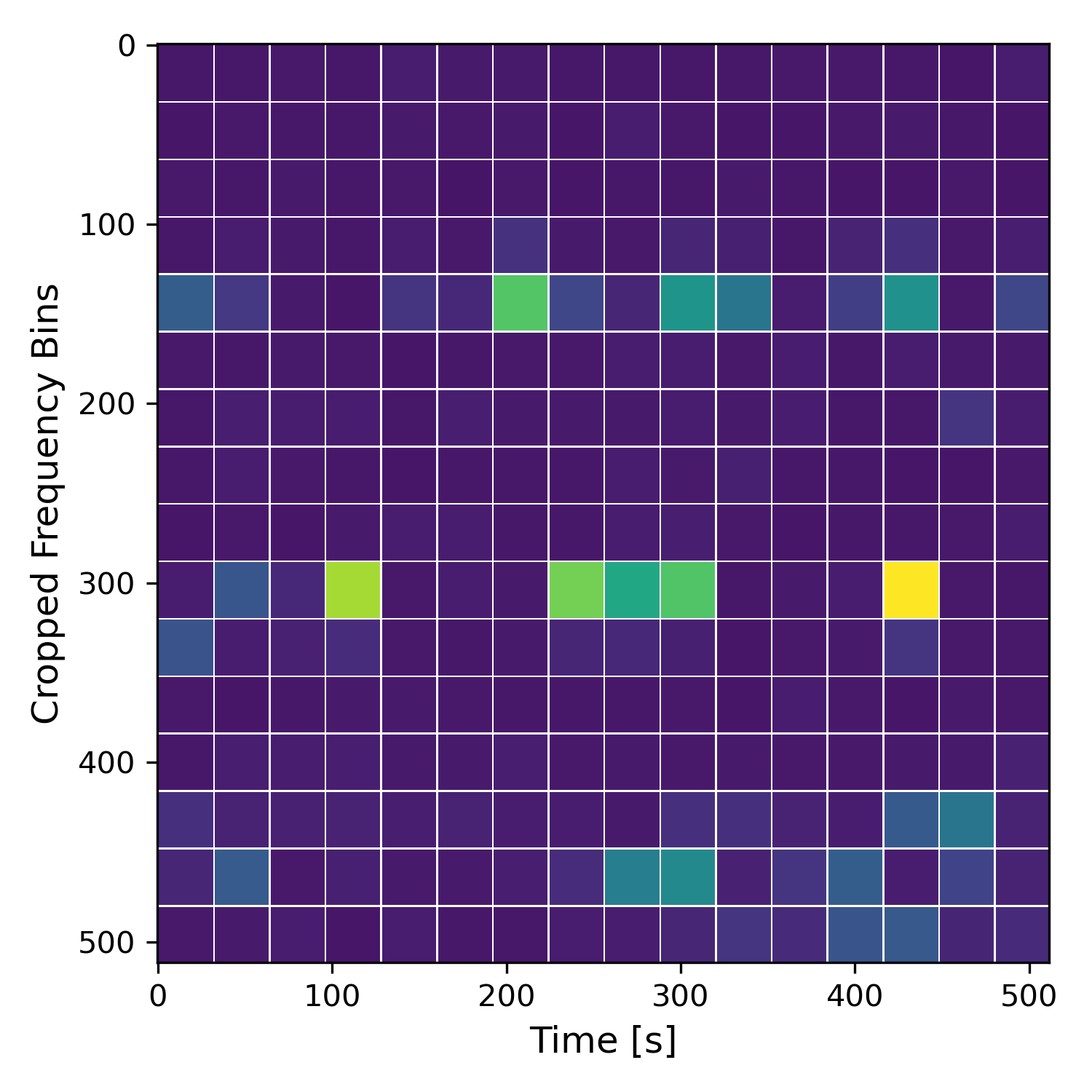}
         \caption{Latent Distances: $z^k$}
         \label{fig:latent_distances}
     \end{subfigure} 
      ~
     \begin{subfigure}[b]{0.32\linewidth}
         \centering
         \includegraphics[width=\linewidth]{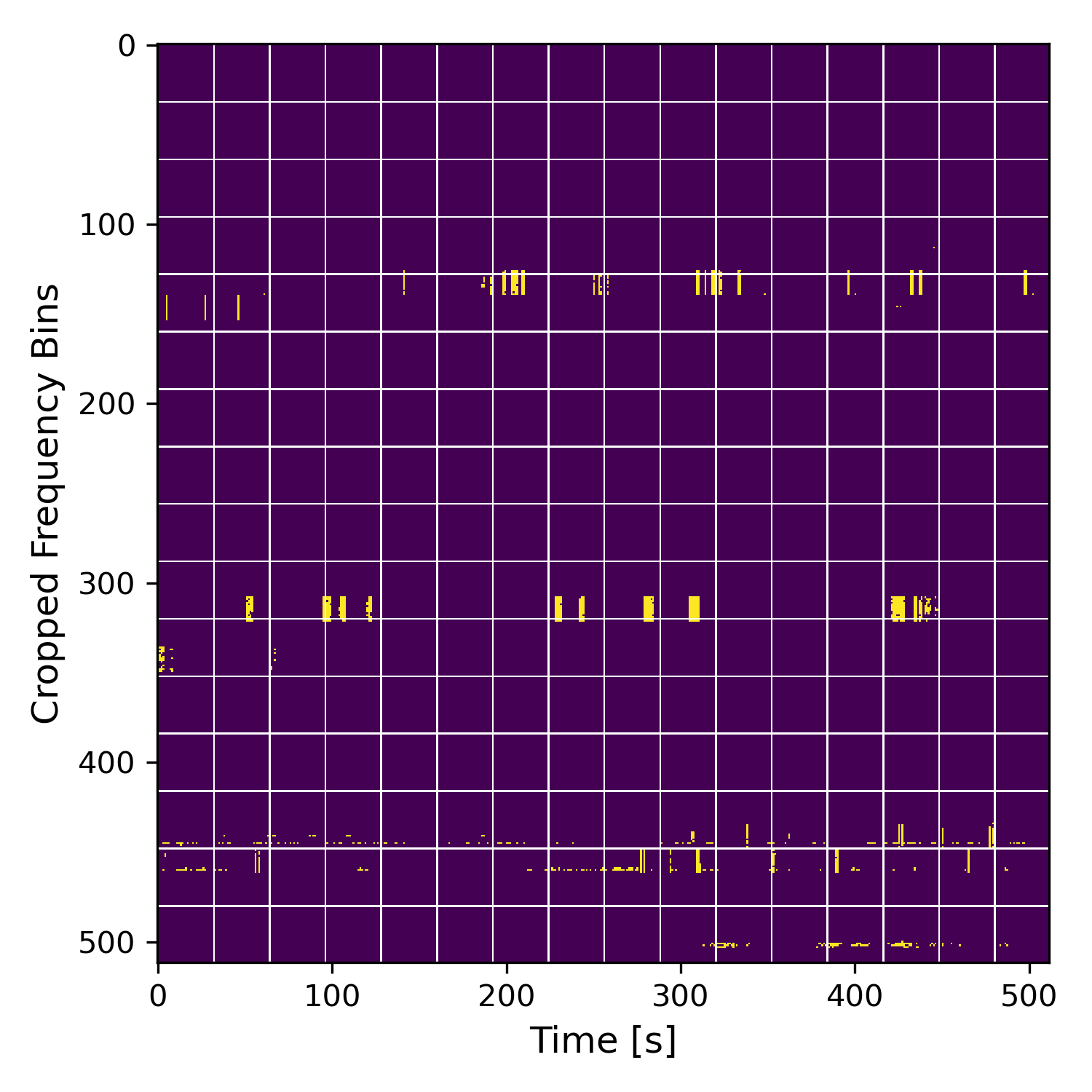}
         \caption{NLN Output: $D_{\text{NLN}}$}
         \label{fig:output}
     \end{subfigure}  
    \caption{Stages of NLN-based RFI detection on the 30th sample of the LOFAR dataset. The white grid illustrates the re-composition of the $32 \times 32$ patches to their corresponding locations in the original spectrogram. Each subplot reflects a part of the modified NLN algorithm from Equation~\ref{eq:NLN}.}
    \label{fig:dataset}
\end{figure*}

Given some trained latent-variable-model such as the discriminative autoencoder, $g(f(V))$, we need to formulate a measure of similarity or difference between the learnt distribution of RFI-free patches and those which are unseen. In practice several options exist such as pixel-level difference~\cite{Akcay}, structured-similarity measure~\cite{Bergmann2019}, residual measures~\cite{Schlegl2017}, purely latent measures~\cite{Bergman2020} and many more. The most important factor when selecting a measure for RFI detection is the resolution of the output. For example, using a purely latent-measure would result in the resolution of the output RFI-masks to be fixed by the resolution of each patch, as shown in Figure~\ref{fig:latent_distances}. However using a pixel-level difference, may cause the predictions to be sensitive to noise. 

To counter-act this problem we propose using a distance function that utilises both latent and pixel-wise measures of difference, namely NLN. 
NLN is a novelty detection technique that works by performing a nearest-neighbour lookup in the latent space of generative models. At training time, it operates as a standard discriminative autoencoder, training on the aforementioned loss.  During inference, a test-sample is given to the model, with the objective to determine which parts (if any) of the input sample are novel. A combination of two metrics is used, the first measures the latent distance from the given test-sample to its nearest neighbours from the distribution of in-lying data, illustrated in Figure~\ref{fig:latent_distances}. The other is the reconstruction error between the given sample and the reconstructions of all its neighbours found in the latent space. Figures~\ref{fig:neighbour_0} and~\ref{fig:neighbour_10} demonstrate that when the autoencoder is trained on only RFI-free data it is capable of only reconstructing the \textit{non-novel} astronomical signals and cannot generate RFI found in the input. In effect, the reconstruction error shown in Figure~\ref{fig:nln_error} has a higher dynamic range than the input.  For more details and analysis of the method see~\cite{Mesarcik2021}.

We modify the original NLN-distance function such that the latent distances are used as a coarse selection for the higher resolution pixel-based error. An illustration of this selection mechanism is shown in Figure~\ref{fig:output}. The modified NLN measure is the reconstruction error of a test-sample's nearest neighbours multiplied by its latent-distance vector, as given by

\begin{equation}
    \label{eq:NLN}
     D_{\text{NLN}} = \big( \dfrac{1}{K} \sum^K |V_{i,n \times n}(\upsilon, \tau,b) - g_{\theta_g}(\mathbf{z}_i^k) | \big) \times D_{\text{latent}}
\end{equation}
where $k$ is the nearest-neighbour of the $i^{\text{th}}$ sample in the latent space given by $\mathbf{z}$.  The nearest-neighbours are selected through the K-Nearest-Neighbours (KNN) algorithm using the default implementation of \texttt{FAISS}~\cite{Johnson2021}. Furthermore, $D_{\text{latent}}$ is the thresholded mean latent-distance vector of the $i^{\text{th}}$ query patch and its $k$ RFI-free neighbours, as given by

\begin{equation}
    D_{\text{latent}} = 
    \begin{cases}
        1,& \dfrac{1}{K} \sum^K |\mathbf{z}_i  - \mathbf{z}_i^k| \geq \mathrm{T} \\
        0,              & \text{otherwise}
    \end{cases}
\end{equation}
We treat both $K$ and $T$ as hyper-parameters of our algorithm and determine them experimentally across our datasets. In effect, the latent distance function offers a coarse resolution view of the RFI, and the reconstruction error offers a finer grained resolution. It must be noted, that the only additional overhead of NLN is that it requires the representations of the training set to be stored.
An illustration of the inference mode of the NLN algorithm is shown in the bottom half of Figure~\ref{fig:diagram}.

\subsection{Architectural considerations}
We use a strided convolutional architecture for the encoder, decoder and discriminator. Both the encoder and decoder have the same architecture except that the decoder uses transposed convolutions in place of the encoder's convolutional layers. Furthermore, the discriminator uses the same architecture as the encoder, except for the final layer, which is a linear layer with a sigmoid activation for the discriminator.     

Several parameters of the architecture are constrained by the chosen patch size and stride width. We find that a patch size of $32\times32$ generally exhibits the best performance, as shown in Figure~\ref{fig:sensitivty}, this limits both the depth and latent dimensionality of our networks. For this reason, the three networks have 2 convolutional layers with $3\times3$ filters and a stride of 2. Each convolutional layer is followed by a batch normalisation layer, and a dropout layer with a rate of 5\% to regularise the network. Lastly, the convolutional output is projected to a specified latent dimensionality by a linear layer.

A base number of filters of 32 is used for the AE and is increased or decreased on each subsequent layer by a factor of 2.  We use \textit{ReLU} activations for all models and they are trained for 100 epochs using ADAM~\cite{Kingma2015} with a learning rate of $1\times10^{-4}$.

\section{Data selection and preprocessing}
\label{sec:Data}
   Existing machine learning based approaches rely on significant amounts of labelled data for training and evaluating the models. By inverting the RFI detection problem, we do not need explicit training labels, but rather rely on the weak labels that typically come without additional cost from data archives such as the LOFAR  Long Term Archive (LTA)~\cite{VanHaarlem2013}. This means that we only need very few expert-labelled examples for evaluation of our model while training on a large dataset as shown in Table~\ref{tab:dataset_statistics}.  
   
   We use two different datasets from two different telescopes to evaluate our work: simulated data from HERA~\cite{DeBoer2017} and calibration data from the LOFAR LTA. We use AOFlagger-based weak-labels for training all models shown in this work on both the HERA and LOFAR datasets. For evaluation with HERA we use the ground truth supplied by the simulator, whereas for the LOFAR we hand-annotate a selection of baselines obtained from the archive. We have made all datasets used in this work publicly available\footnote{\href{https://zenodo.org/record/6724065}{https://zenodo.org/record/6724065}}.   

\begin{figure*}
     \centering
     \begin{subfigure}[b]{0.32\linewidth}
         \centering
         \includegraphics[width=\linewidth]{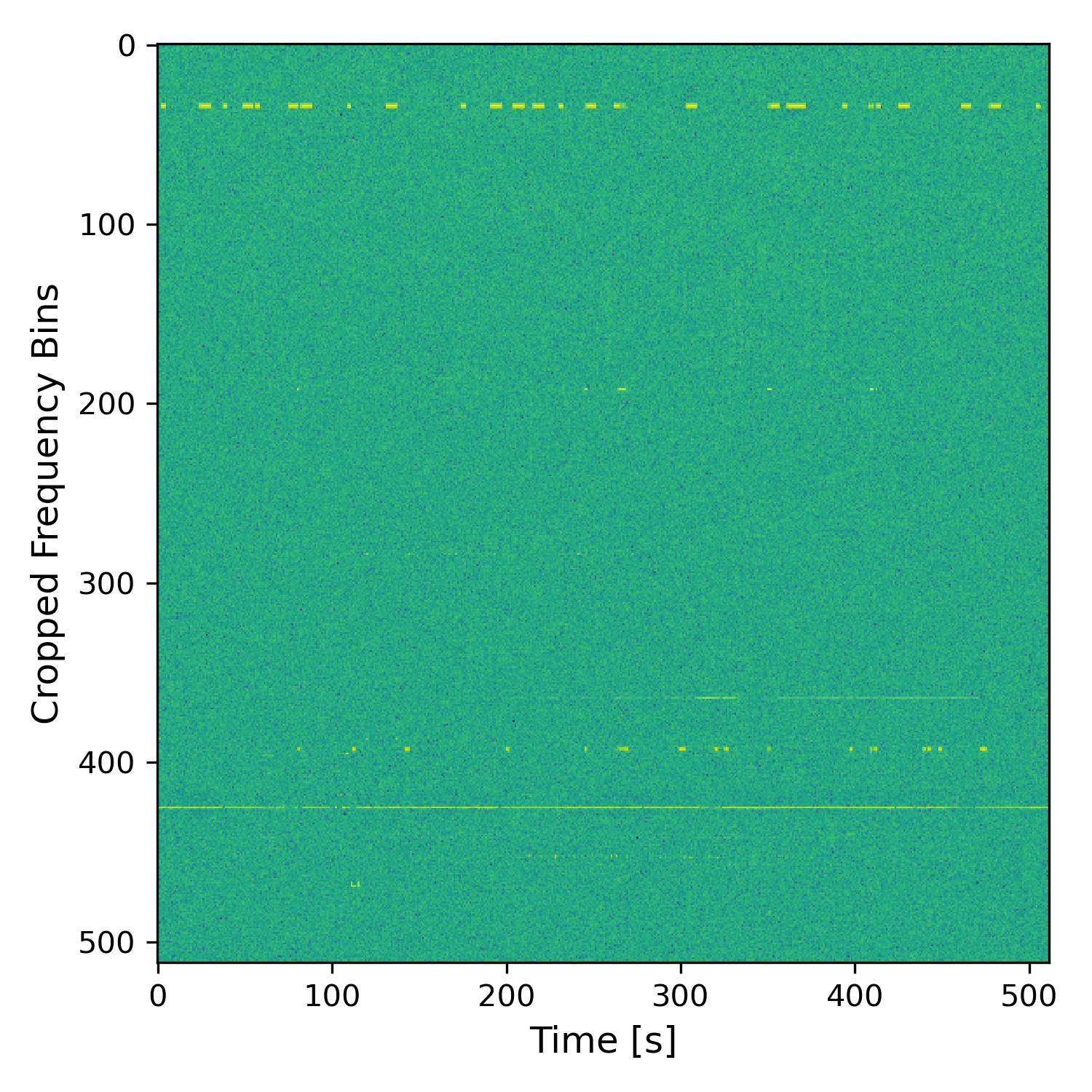}
         \caption{Log-magnitude spectrum for 18$\text{th}$ sample}
         \label{fig:data}
     \end{subfigure}
     ~~
     \begin{subfigure}[b]{0.32\linewidth}
         \centering
         \includegraphics[clip,width=\linewidth]{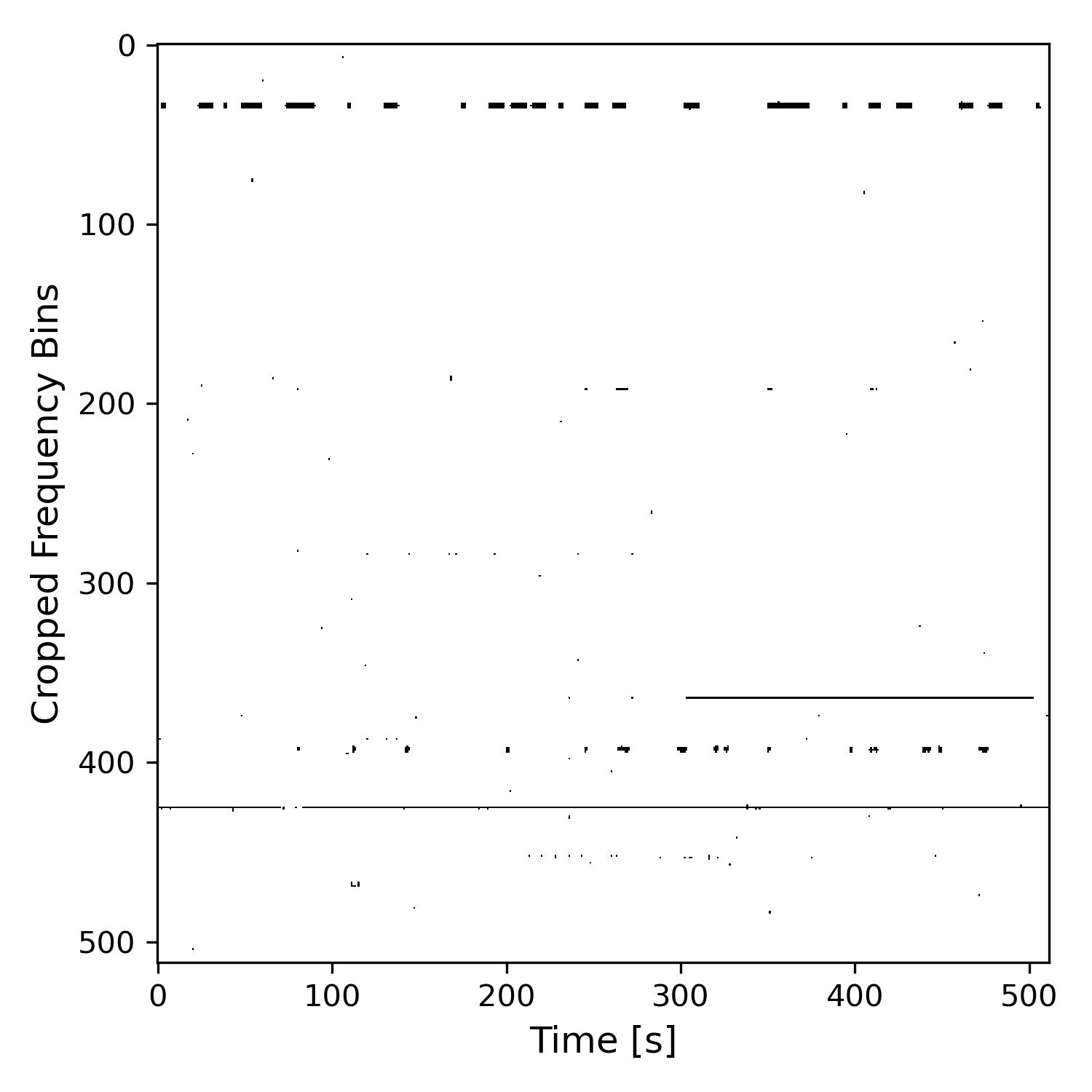}
         \caption{AOFlagger annotation for 18$\text{th}$ sample}
         \label{fig:aoflag}
     \end{subfigure}
     ~~
     \begin{subfigure}[b]{0.32\linewidth}
         \centering
         \includegraphics[width=\linewidth]{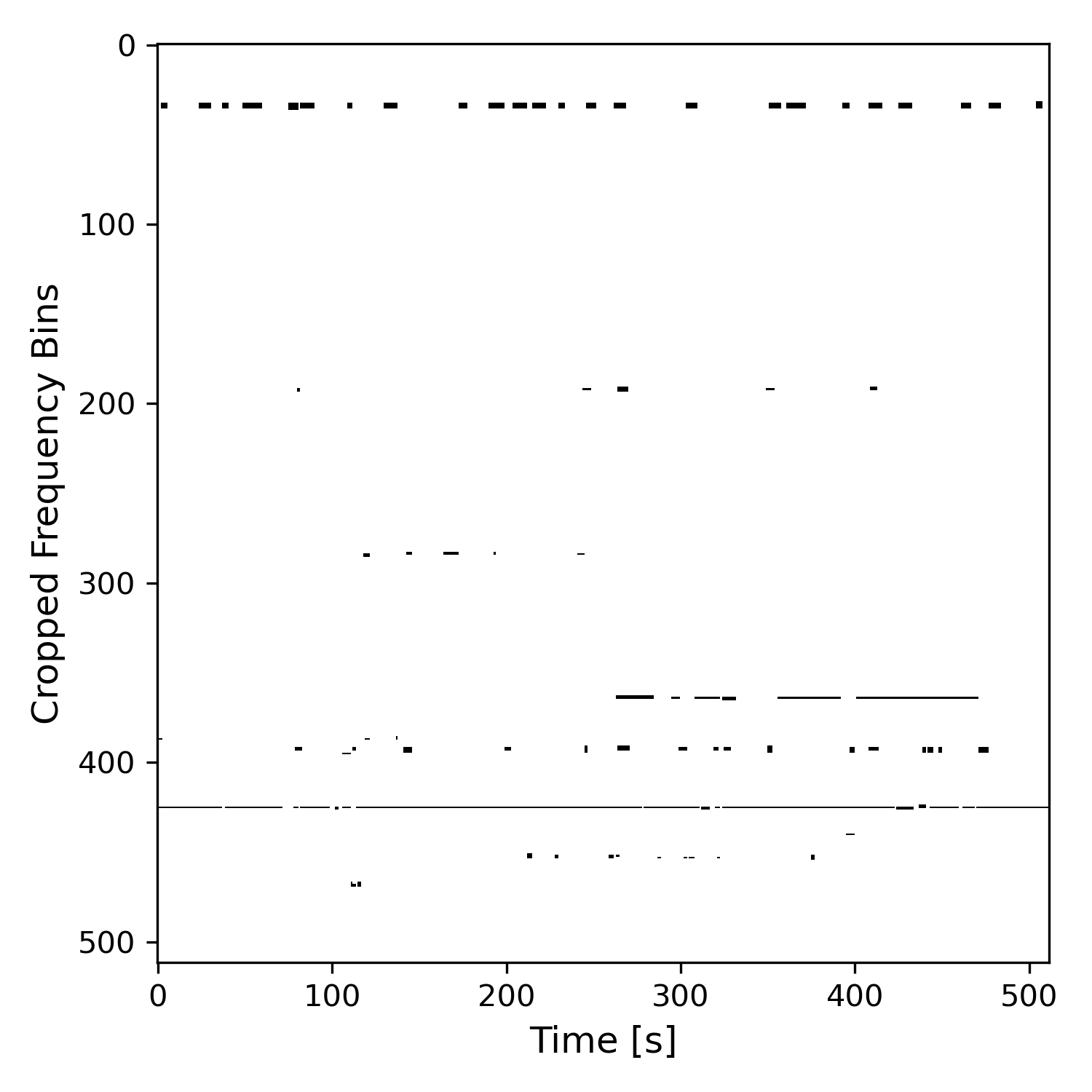}
         \caption{Expert-labelled annotation for 18$\text{th}$ sample}
         \label{fig:labelme}
     \end{subfigure}
     \hfill
     \begin{subfigure}[b]{0.32\linewidth}
         \centering
         \includegraphics[width=\linewidth]{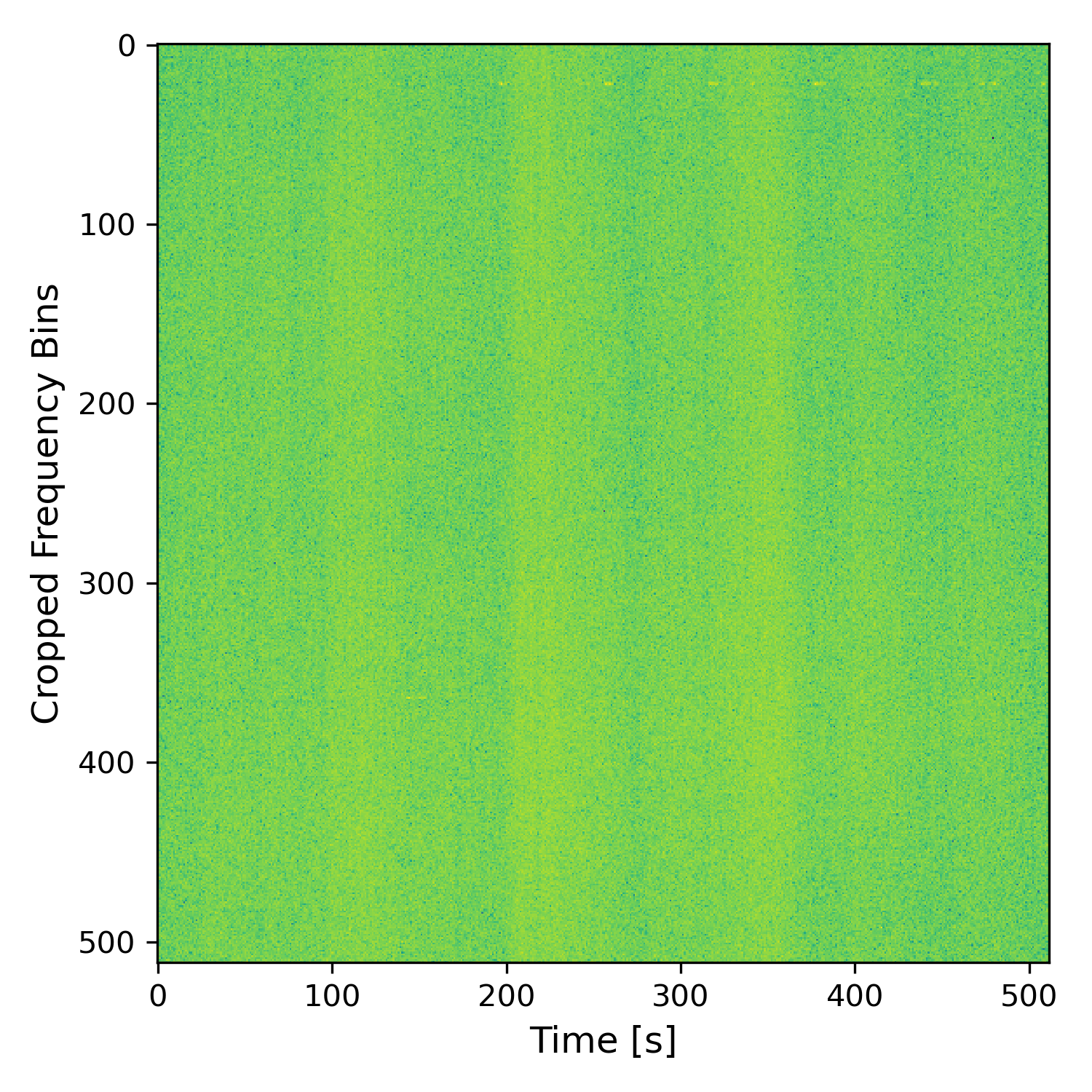}
         \caption{Log-magnitude spectrum for 47$\text{th}$ sample}
         \label{fig:data}
     \end{subfigure}
     ~~
     \begin{subfigure}[b]{0.32\linewidth}
         \centering
         \includegraphics[width=\linewidth]{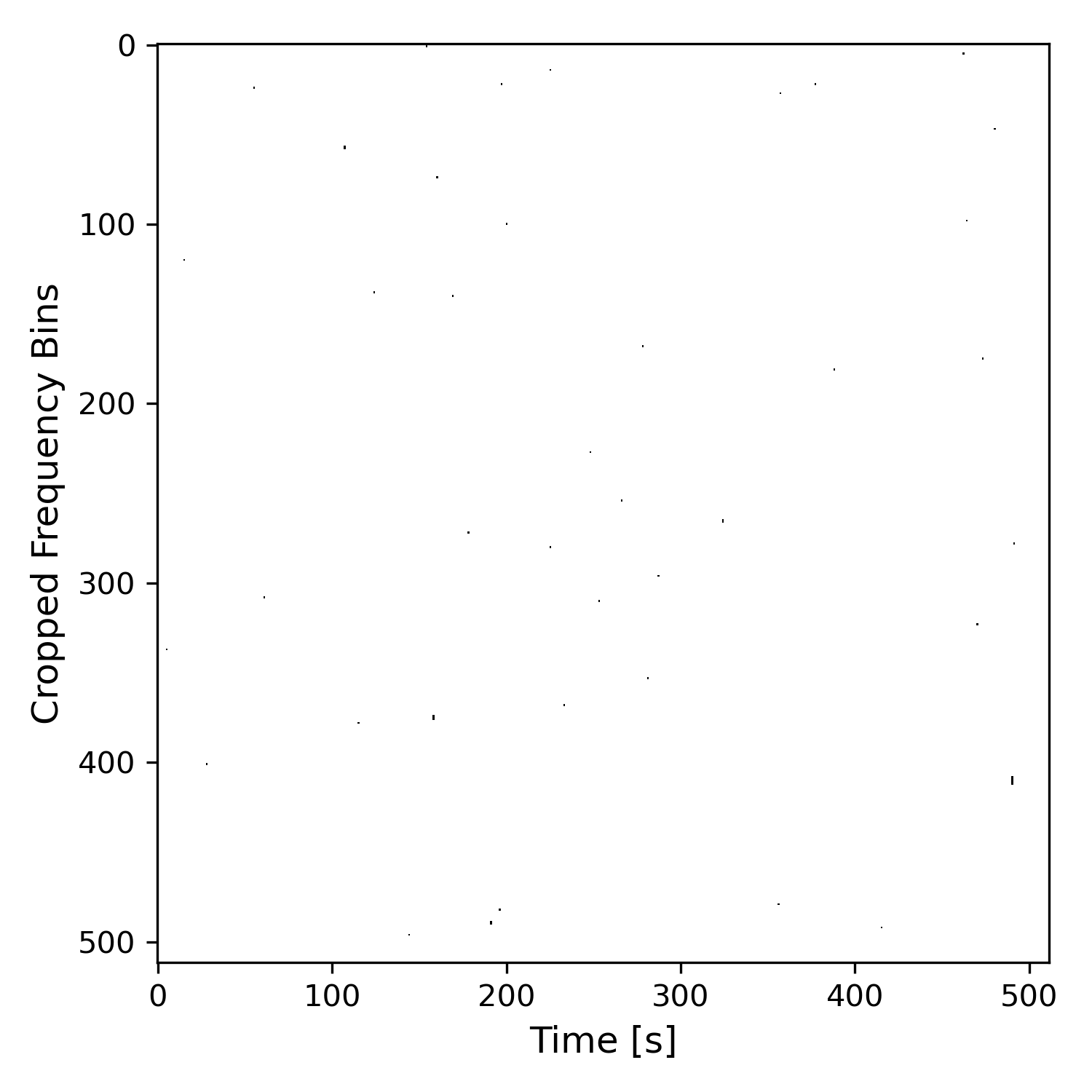}
         \caption{AOFlagger annotation for 47$\text{th}$ sample}
         \label{fig:aoflag}
     \end{subfigure}
     ~~
     \begin{subfigure}[b]{0.32\linewidth}
         \centering
         \includegraphics[width=\linewidth]{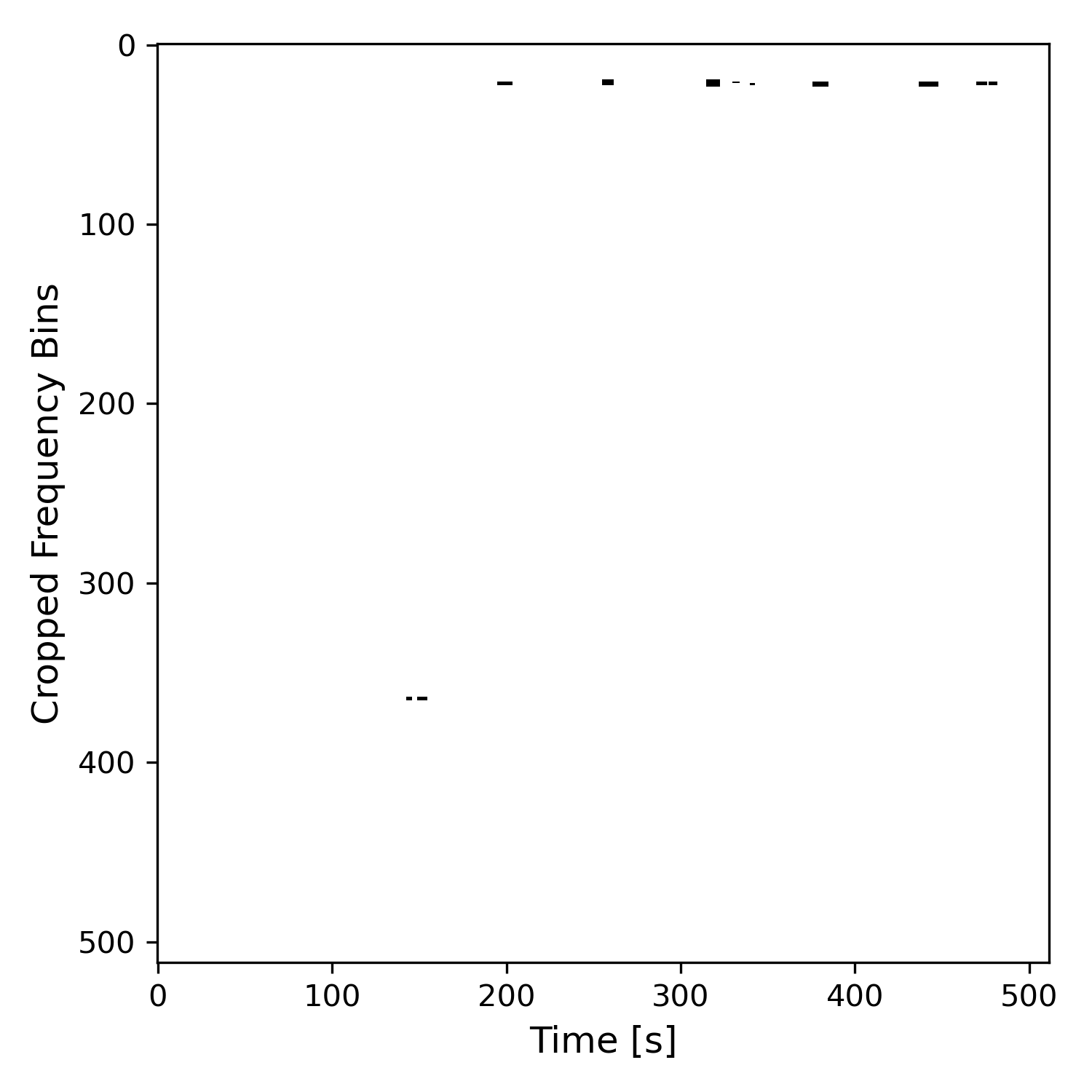}
         \caption{Expert-labelled annotation for 47$\text{th}$ sample}
         \label{fig:labelme}
     \end{subfigure}
        \caption{Spectrograms and their masks from the LOFAR dataset with the highest overlap between AOFlagger and expert-labelled annotations (top row) and the lowest overlap (bottom row), where similarity is measured using the F1-score.}
        \label{fig:dataset}
\end{figure*}
\subsection{Simulated HERA dataset}
    The HERA simulator~\footnote{\href{hera\_sim}{https://github.com/HERA-Team/hera\_sim}} generates complex spectrograms from a simulated radio telescope. It uses models of diffuse sources, multiple types of RFI emissions and systematic models of the HERA telescope for parameters such as antenna cross-coupling, band-pass effects and more. Importantly for this work, the simulator gives operators a fine-grained control over the generated RFI types as well as their pixel-precise ground truth maps. Thanks to these properties, we can use the HERA data for the validation of our approach.
    
    In this work, we simulate a hexagonal array with 14.6~m between each station, as specified by the 19 element array~\cite{DeBoer2017}. To synthesise our dataset, we create a 30 minute observation with an integration time of 3.52~s and bandwidth of 90~MHz from 105~MHz to 195~MHz (with 512 frequency channels). The specific integration interval is used to ensure that the resulting spectrograms are square, this is done to simplify the arithmetic of creating and reconstructing the spectrograms from patches. Furthermore, we use the default number of diffuse galactic emissions specified in~\cite{Kerrigan2019} for the \texttt{H1C} observation season from 2017 to 2018. Finally, we include additive thermal noise 180~K at 180~MHz. 
   
   \begin{table}
    \centering
    \resizebox{\linewidth}{!}{
    \begin{tabular}{|c|c|c|c|c|c|}
         \hline
         \textbf{Dataset} & \textbf{\# Baselines}&\textbf{ \# Training samples}& \textbf{\# Test Samples}&\textbf{\% RFI} \\
         \hline
         \hline
         \texttt{HERA} & 28 & 420 & 140 & 2.76 \\
         \hline
         \texttt{LOFAR} & 2775 & 7500 & 109 & 1.26  \\
         \hline
    \end{tabular}}
    \caption{Attributes of each dataset used for training and evaluation.  The low test-train-ratio is due to the use of weak-labels generated from classical methods, that are not directly used in the NLN training.}
    \label{tab:dataset_statistics}
\end{table}   
   
    In the generation of our training set we individually synthesise multiple RFI emissions based on the models specified in~\cite{Kerrigan2019}. These being narrow-band continuous emissions modelled satellite communication such as \textit{ORBCOM}, broad-band transient emissions that imitate events such as lightning,  as well as narrow-band burst RFI based on ground communication. Additionally, we include random single time-frequency \textit{blips}.
  
    Using the hexagonal array we simulate 28 baselines of both auto and cross correlations. We repeat this 20 times to obtain 560 complex spectrograms to train and evaluate our models with. This approach was deemed more appropriate than simulating a single long observation as performed in~\cite{Kerrigan2019}, as there is more diversity in the RFI landscape from multiple initialisations of the simulator. The simulated data has an RFI occupancy rate of 2.76\% and is split into 75\% for training and 25\% for testing as reported in Table~\ref{tab:dataset_statistics}.
    
    \begin{figure*}
        \centering
        \includegraphics[trim={0.3cm 0.2cm 0.3cm 0.35cm} ,clip,width=\linewidth]{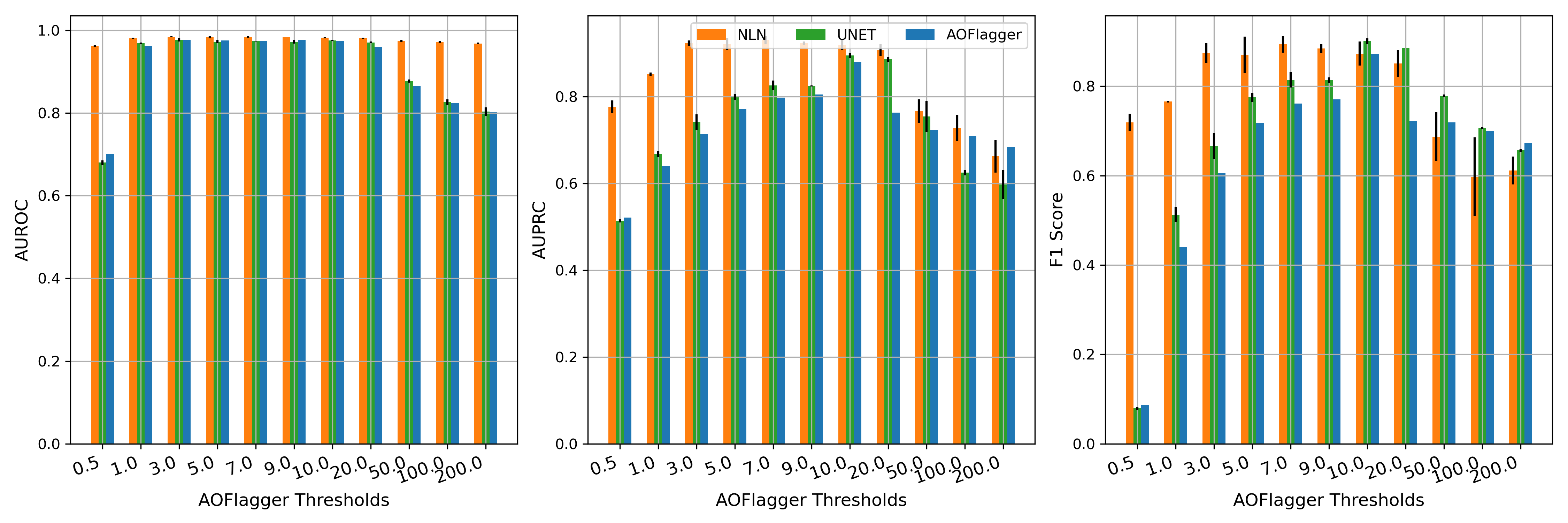}
        \caption{Performance of NLN, UNET and AOFlagger on the HERA data. Both U-Net and NLN are trained on the AOFlagger masks at for given starting threshold and evaluated on the simulator-ground truth. } 
        \label{fig:HERA_Thresholds}
    \end{figure*}

    The simulated data is preprocessed before training and evaluation. For simplicity purposes, we only use the magnitude of the complex visiblities. To deal with the high dynamic range, we clip the data between $[|\mu - \sigma|,\: \mu + 4\sigma]$ and take the natural $\log$ of the clipped spectrograms. Finally before training we standardise our data between 0 and 1 to ensure the gradients while training do not explode or vanish.

\subsection{LOFAR}
    As previously mentioned we use publicly available data from the LOFAR Long Term Archive (LTA)~\cite{VanHaarlem2013}. Five measurements were randomly selected from 2017 to 2018 for evaluation our model. We select calibration measurements that point at strong radio sources, using 51 stations in the band 120-190~MHz for 600s. The precise details of the observations are available at \cite{Mesarcik2022}.
    
    A common challenge in applying machine learning to radio astronomy is the amount of data generated by observations. This is especially problematic when training models with limited GPU memory. For example the 5 raw Measurement Sets (MS) used in this work are 1.7~TB in size. Therefore, in order to decrease the data-size, we use only the magnitude of the first stokes parameter and randomly sample 1500 baselines of each observation. This effectively reduces the dataset size to $\sim$10~GB.  
    
    For purposes of further reducing the training memory footprint of the data and simplifying the arithmetic of reconstructing the spectrograms from their respective patches, we first create approximately square spectrograms. As there are 599 time-samples per observation and 16 subbands per channel, we concatenate each 44 consecutive subbands together to form spectrograms size of 
    $599\times 616$. Additionally, we discard the first and last channel of each subband due to bandpass-effects. Finally we randomly crop the resulting spectrograms to $512\times 512$. This is done, as cropping gives an equal representation of all frequency bands. 
    
    For the evaluation of the models on the LOFAR dataset, we randomly select 109 baselines for expert labelling. This is in line with the number of baselines used for evaluation in the simulator-based setting. As noted in~\cite{Akeret2017a, Kerrigan2019, Sadr2020}, there are often discrepancies between the AOFlagger masks and those given by an expert, this is highlighted in Figure~\ref{fig:dataset}. For validation and evaluation of this work we treat the hand-annotated examples as the ground truth, as described in Section~\ref{sec:Results}.  Furthermore, we use the AOFlagger masks associated with the measurement sets from the LTA for training our models.

    In Table~\ref{tab:dataset_statistics} we report the percentage of RFI contamination and the dataset sizes. We ensured that the contamination is in line with what is reported in~\cite{VanHaarlem2013}. Furthermore, as our method does not rely on human-labelled examples to train, we only require a small number of expert-labelled examples to evaluate the performance of our models.

    We pre-process the LOFAR data in a similar manner to the simulated data. We first clip the data between $[|\mu - \sigma|,\: \mu + 20\sigma]$ and followed by the natural logarithm. We finally standardise the data between 0 and 1 to ensure the gradients while training do not explode or vanish.  
\section{Results}
\label{sec:Results}
We evaluate the performance of NLN applied to RFI detection experimentally on both simulated and real data from HERA and LOFAR respectively. Unlike previous works, we perform our evaluation in a two-step process, first by finding an appropriate AOFlagger threshold to generate the training annotations and then calculating the respective performance of each model on the real ground truth. To ensure the correctness of the evaluation we use the ground-truth masks from the simulator in the test-set for the HERA models and human-generated annotations for the testing of the LOFAR models.   

\subsection{Evaluation Methodology}
Following previous efforts to evaluate the performance of RFI detection~\cite{Offringa2012,Akeret2017, Kerrigan2019, Sadr2020}, we use the three most common metrics: the Area Under the Receiver Operating Characteristic (AUROC) score, Area Under Precision Recall Curve (AUPRC) and F1-Score. The AUROC metric evaluates the ratio of True Positive Rate (TPR) and False Positive Rate (FPR) across several thresholds. In this case, the TPR is the fraction of RFI that is correctly classified as RFI and the FPR is the fraction of misclassified signals. The AUPRC metric gives the ratio of precision and recall when the model's output is evaluated across several thresholds. In this case precision refers to the fraction of correctly classified RFI across all RFI predictions, and the recall is simply the TPR. Finally, the F1 score is the harmonic mean of the precision and recall for a given threshold. For all evaluations across all models in this work the threshold is fixed to the maximum obtainable F1 score.  

In the class-imbalanced scenario of RFI detection, high AUROC scores imply that a model is effective in classifying the majority class. This means that all non-RFI signals are detected as not RFI. Conversely, the AUPRC and F1 scores focus on the minority class, meaning that when AUPRC is high, the model is better at detecting RFI with a low RFI misdetection rate. Therefore, in order to maintain consistency with previous works' evaluations and to give insight into a model's performance on the both the majority and minority classes, we evaluate using both AUPRC and AUROC.

We use the AOFlagger masks to train all models in this work. In the case of LOFAR, we use the flags provided by the \texttt{FLAG} field of the measurement sets obtained from the LTA. For the HERA dataset, the optimal flagging strategy is determined experimentally. As there is no pre-specified strategy for the HERA telescope, we test the HERA dataset on all available strategies for several different base-thresholds. We find that the \texttt{bighorns}-telescope strategy with a starting threshold of 10 to be optimal with respect to the joint-maximum of AOFlagger across AUROC, AUPRC and F1-score as shown in Figure~\ref{fig:HERA_Thresholds}.

For comparison with existing work we select the state-of-the-art RFI detection models. As described in Section~\ref{sec:Related_Work}, we evaluate with supervised-segmentation algorithms, based off the U-Net~\cite{Ronneberger2015} architecture. These being, the magnitude-only U-Net for RFI detection~\cite{Akeret2017,Kerrigan2019} as well as its residual adaptions, R-Net~\cite{Sadr2020} and RFI-Net~\cite{Yang2020}. Additionally, we measure the AOFlagger on both datasets and report its performance. We train and evaluate every model 3 times with a randomly initialised seed and report the mean and standard deviation for each evaluation.  

For both datasets, we perform an independent coarse grid search across the hyper-parameters of NLN. These being patch size ($n$), the number of latent space dimensions ($L$), number of neighbours ($K$), maximum latent distance threshold ($T$) and discriminative training weight ($\alpha$). We determine the optimal hyper-parameters based on the average maximum of AUROC and AUPRC. For LOFAR and HERA we use 16 and 20 neighbours respectively and find a patch size of $32\times 32$ to be optimal for both. Furthermore, the latent dimension size of 8 is used for HERA whereas we find 32 dimensions to be optimal for LOFAR. We threshold the latent distance vector at its $66^{th}$ percentile for LOFAR and the $10^{th}$ percentile for HERA. Finally, we find that the optimal discriminative hyper-parameter $\alpha$ is $0.6$ for both the HERA and LOFAR datasets. For all other models, we use all parameters specified by the authors other than patch size, which we fix to $32 \times 32$ in order to keep comparison consistent.

\begin{table}
    \centering  
    \resizebox{\linewidth}{!}{
    \begin{tabular}{|c|c|c|c|}
         \hline
         \textbf{Model} & \textbf{AUROC} &  \textbf{AUPRC}& \textbf{F1-Score}\\
         \hline
         \hline
         AOFlagger~\cite{Offringa2012a}& 0.9736 & 0.8799 & 0.8729 \\
         \hline
         U-Net~\cite{Akeret2017a} & 0.9746 $\pm$ 0.0009 & 0.8963 $\pm$ 0.0074 & 0.9015 $\pm$ 0.0074\\
         \hline
         RFI-Net~\cite{Yang2020} & 0.9728  $\pm$ 0.0007 & 0.8898 $\pm$ 0.0086 & 0.8950 $\pm$ 0.0068\\
         \hline
         RNet~\cite{Sadr2020} & 0.9751 $\pm$ 0.0016 & 0.8456 $\pm$ 0.0134 & 0.8460  $\pm$ 0.0223 \\
         \hline
         NLN (ours) & \textbf{0.9805 $\pm$  0.0017}  & \textbf{0.9265 $\pm$ 0.0117} & \textbf{0.9103 $\pm$ 0.0229}\\
         \hline
         \end{tabular}}
    \caption{Performance of RFI detection models on the simulated HERA dataset when trained using the AOFlagger annotation at a threshold of 10 and evaluated on the ground truth from the simulator. We do not report standard deviation of AOFlagger as it is deterministic. Best scores in bold.}
    \label{tab:HERA_results}
\end{table}

The difference in the hyper-parameters between the datasets is due to the increased complexity of the real LOFAR data relative to the simulated HERA data. This complexity is due to stochastic instrumentation effects, ionospheric artefacts, increased dynamic range and many more. Therefore, the autoencoder requires a larger latent dimensionality to better represent this increased complexity. Similarly, an increased latent distance threshold is used to mitigate the elevated reconstruction error-based noise.

Finally, to validate the suitability of the NLN algorithm for novelty-based RFI detection we compare it against three commonly used novelty detection techniques. In this case we consider Deep K-Nearest Neighbours (DKNN)~\cite{Bergman2020}, autoencoding models with an L2 loss~\cite{Bergmann2019a} and SSIM-based loss ~\cite{Bergmann2019}. 

\begin{figure}
    \centering
    \includegraphics[trim={0.1cm 0.4cm 0.3cm 0.35cm} ,clip,width=\linewidth]{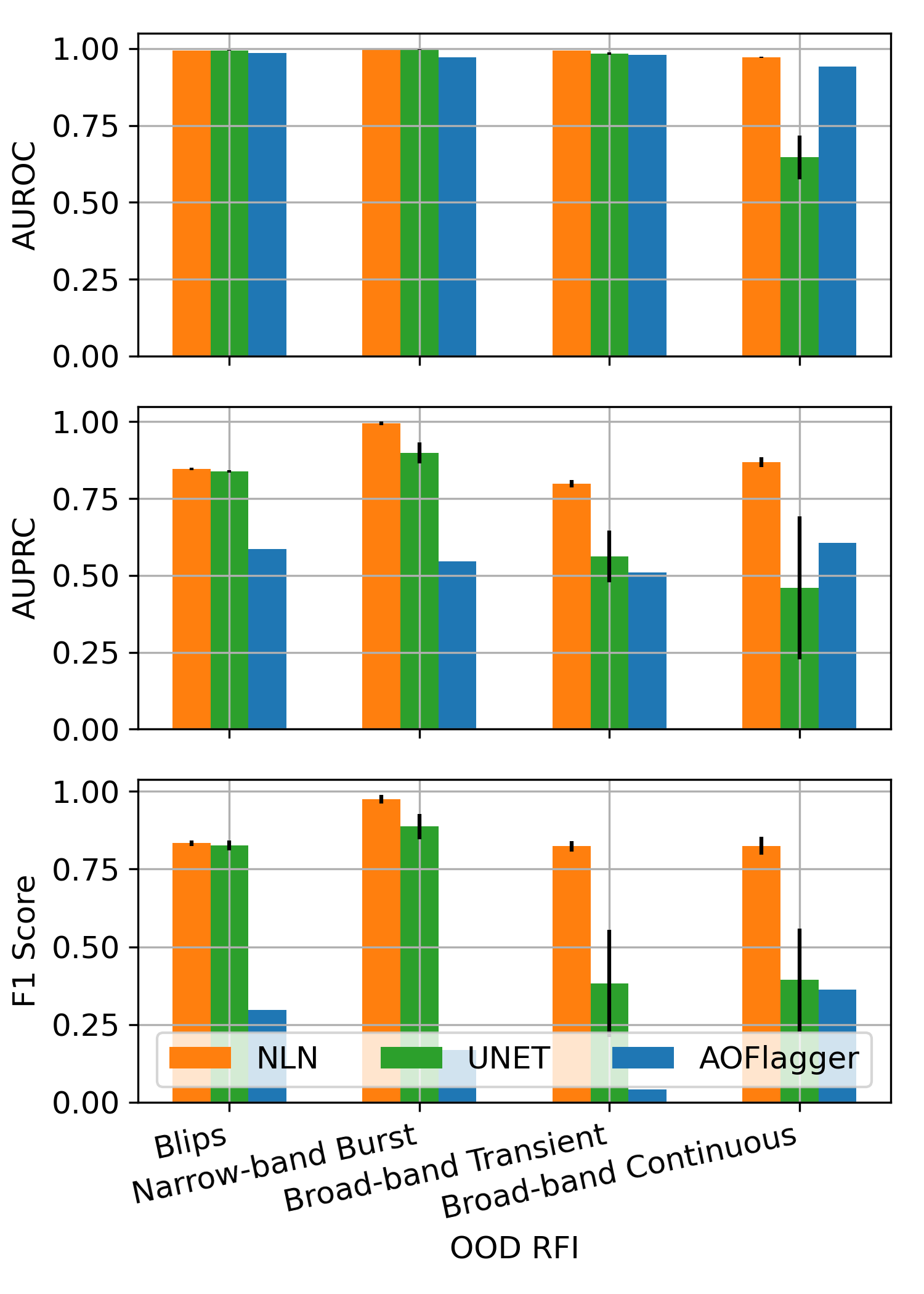}
    \caption{Out-Of-Distribution (OOD) RFI detection performance of the NLN, U-NET and the AOFlagger. As the RFI is excluded in the training process of NLN, the method is unaffected by OOD effects. }
    \label{fig:HERA_OOD}
\end{figure}

\subsection{HERA Results}
\label{sec:HERA_results}
In Figure~\ref{fig:HERA_Thresholds} we illustrate the performance sensitivity of the magnitude-based U-Net and NLN when modifying the AOFlagger starting threshold for the \texttt{bighorns}-strategy. It is clear NLN is less sensitive to changes in the validity of the training data, exhibiting little variation relative to the AOFlagger-based masks. This is because the model is not directly trained on the AOFlagger-based masks. However UNET's performance is shown to be more dependent on the accuracy of the AOFlagger masks, indicating that it indeed is over-fitting to the training labels. 

For low thresholds ($>5$) the training data is over-flagged, meaning that all RFI is flagged but large percentage of the astronomical data is as well. Conversely, for high thresholds ($<10$) the data is under-flagged, meaning that some RFI is not flagged. From this intuition, it is clear that the NLN algorithm is less sensitive to over-flagging, but its performance deteriorates when the training data is under-flagged. For the experiments using the simulated HERA dataset in the remainder of this paper, we fix the AOFlagger threshold to 10, as it gives optimal flagging performance. 

In Table~\ref{tab:HERA_results} we show the performance of NLN for RFI detection relative to the current state-of-the-art. It is clear that the NLN offers superior performance across all metrics demonstrating the success of our approach to use clean data for training.

In Figure~\ref{fig:HERA_OOD} we demonstrate an interesting unintended consequence of our work in the HERA setting. Here, we train each model on the dataset but excluding a particular type of RFI, then during testing we expose the model to the unseen RFI type. This paradigm effectively evaluates how well models generalise to Out-Of-Distribution (OOD) RFI. As the training process of NLN excludes all RFI (OOD or not) from the training set, it does not learn explicit models of the RFI, in effect our method significantly outperforms both supervised and classical methods across all metrics. This is also clearly demonstrated by the variance across each experiment.

\begin{table}
    \centering  
    \resizebox{\linewidth}{!}{
    \begin{tabular}{|c|c|c|c|}
         \hline
         \textbf{Model} & \textbf{AUROC} &  \textbf{AUPRC}& \textbf{F1-Score}\\
         \hline
         \hline
         AOFlagger~\cite{Offringa2012a} & 0.7883 & 0.5716 & 0.5698 \\
         \hline
         \hline
         U-Net~\cite{Akeret2017a} &  0.8017 $\pm$  0.0058 & 0.5920 $\pm$ 0.0031 & 0.5876  0.0031\\
         \hline
         RFI-Net~\cite{Yang2020} & 0.8109 $\pm$  0.0037 & 0.5991 $\pm$ 0.0038 & \textbf{0.5979 $\pm$ 0.0012}\\
         \hline
         RNet~\cite{Sadr2020} & 0.8301 $\pm$  0.0084 & 0.5495 $\pm$ 0.0145 &  0.5286 $\pm$ 0.0195 \\
         \hline
         \hline
         AE-L2~\cite{Bergmann2019a} & 0.8397 $\pm$ 0.0019  &   0.3933 $\pm$ 0.0036 & 0.4491 $\pm$ 0.0007 \\
         \hline
         AE-SSIM~\cite{Bergmann2019} & 0.7748 $\pm$ 0.0046 & 0.3913 $\pm$ 0.0186 & 0.4801 $\pm$ 0.0115 \\
         \hline
         DKNN~\cite{Bergman2020} & 0.8285  &  0.0704  &  0.1528  \\
         \hline
         \hline
         NLN (ours) & \textbf{0.8622 $\pm$   0.0006} & \textbf{0.6216 $\pm$ 0.0005} & 0.5114 $\pm$ 0.0004\\
         \hline
         \end{tabular}}
    \caption{Performance of RFI detection models on the real LOFAR dataset when trained using the AOFlagger annotations from the LTA and evaluated on the expert-labelled ground truth. We do not report AOFlagger and DKNN standard deviations as they are deterministic.   Best scores in bold.}
    \label{tab:LOFAR_results}
\end{table}

\subsection{LOFAR results}
\label{sec:LOFAR_results}
\begin{figure*}
     \centering
     \begin{subfigure}[b]{0.40\linewidth}
         \centering
         \includegraphics[trim={0.1cm 0.1cm 0.1cm 0.1cm},clip,width=0.9\linewidth]{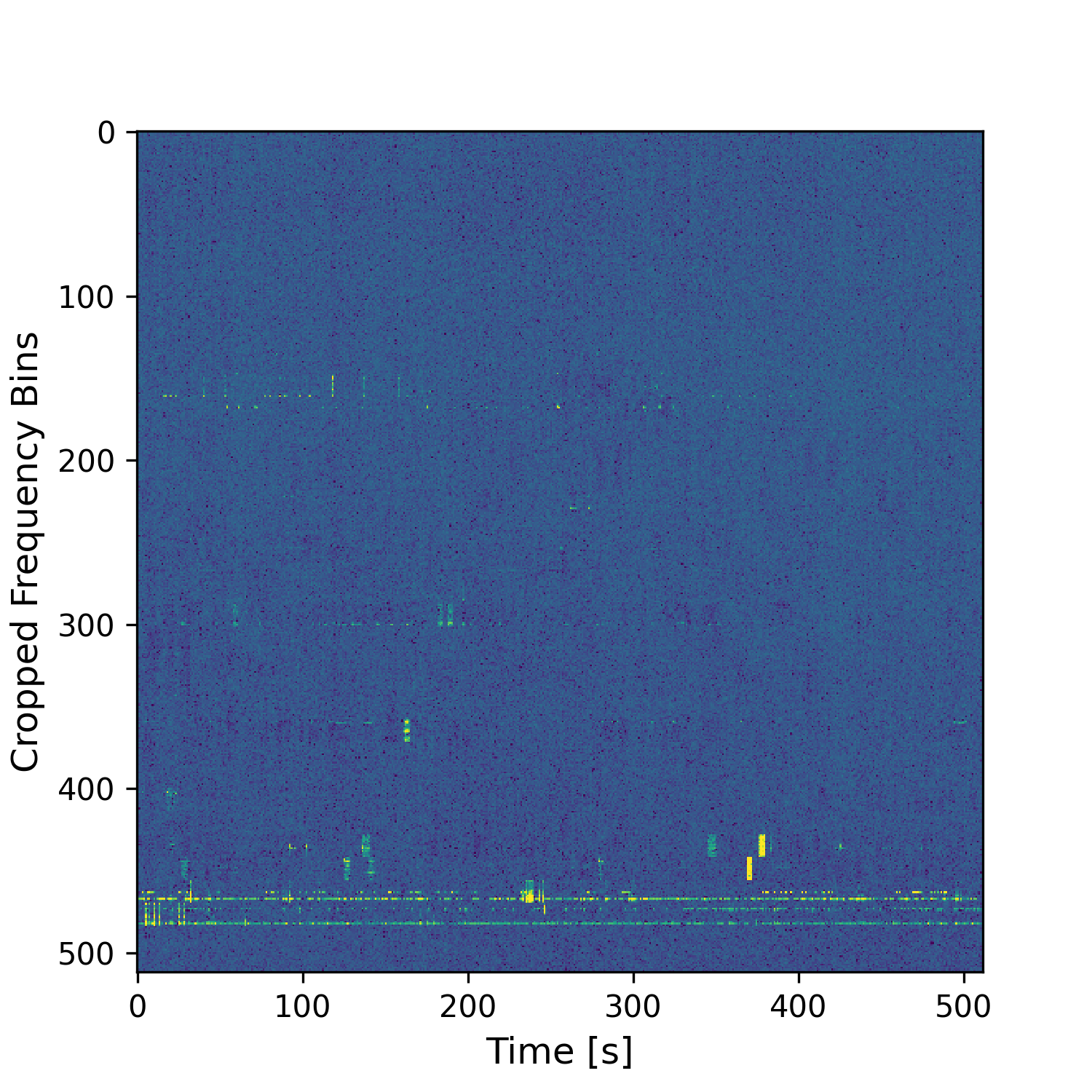}
         \caption{Input}
         \label{fig:95_input}
     \end{subfigure}
     ~~
     \begin{subfigure}[b]{0.40\linewidth}
         \centering
         \includegraphics[trim={0.1cm 0.1cm 0.1cm 0.1cm},clip,width=0.9\linewidth]{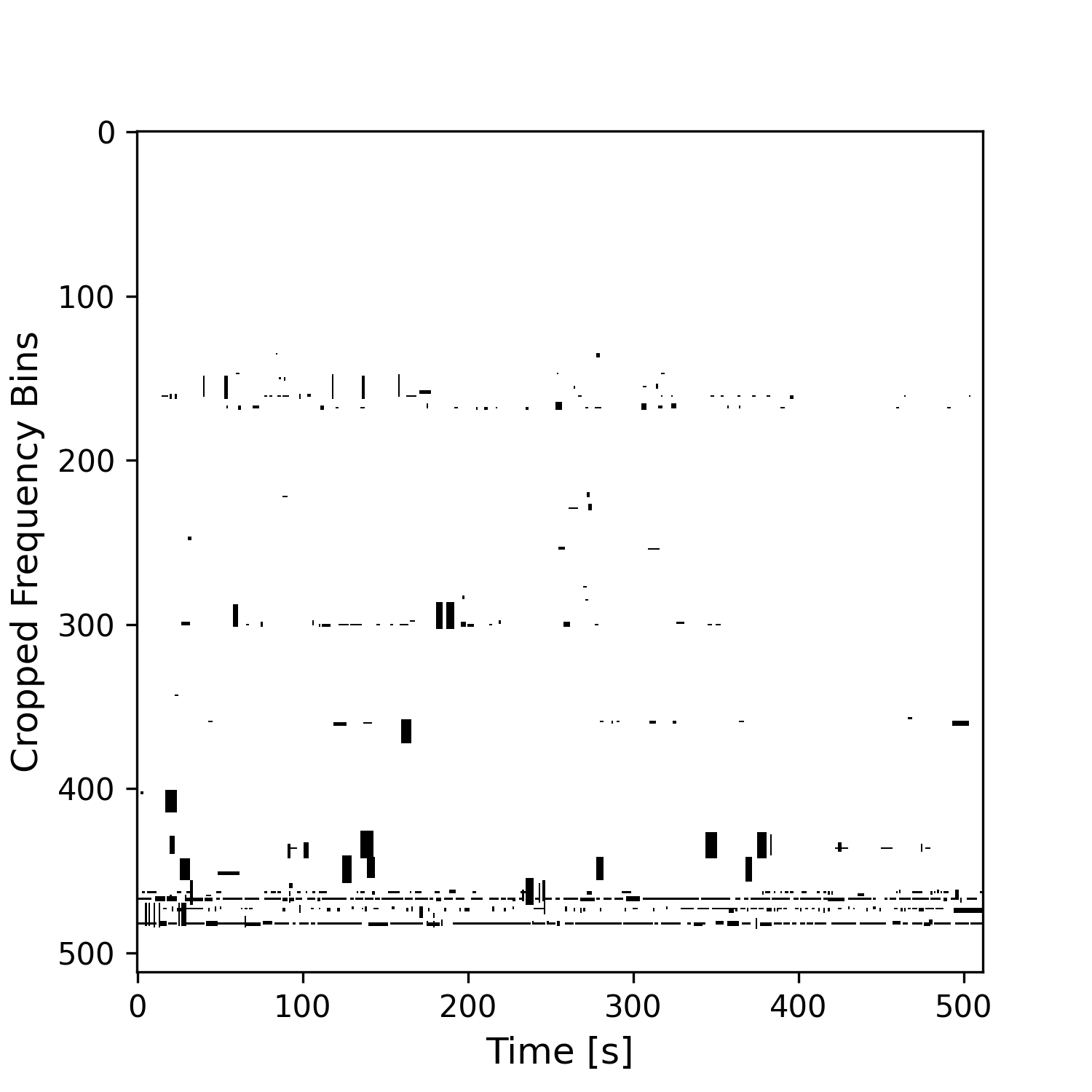}
         \caption{Mask}
         \label{fig:95_masks}
     \end{subfigure}
     \\
     \begin{subfigure}[b]{0.40\linewidth}
         \centering
         \includegraphics[trim={0.1cm 0.1cm 0.1cm 0.1cm},clip,width=0.9\linewidth]{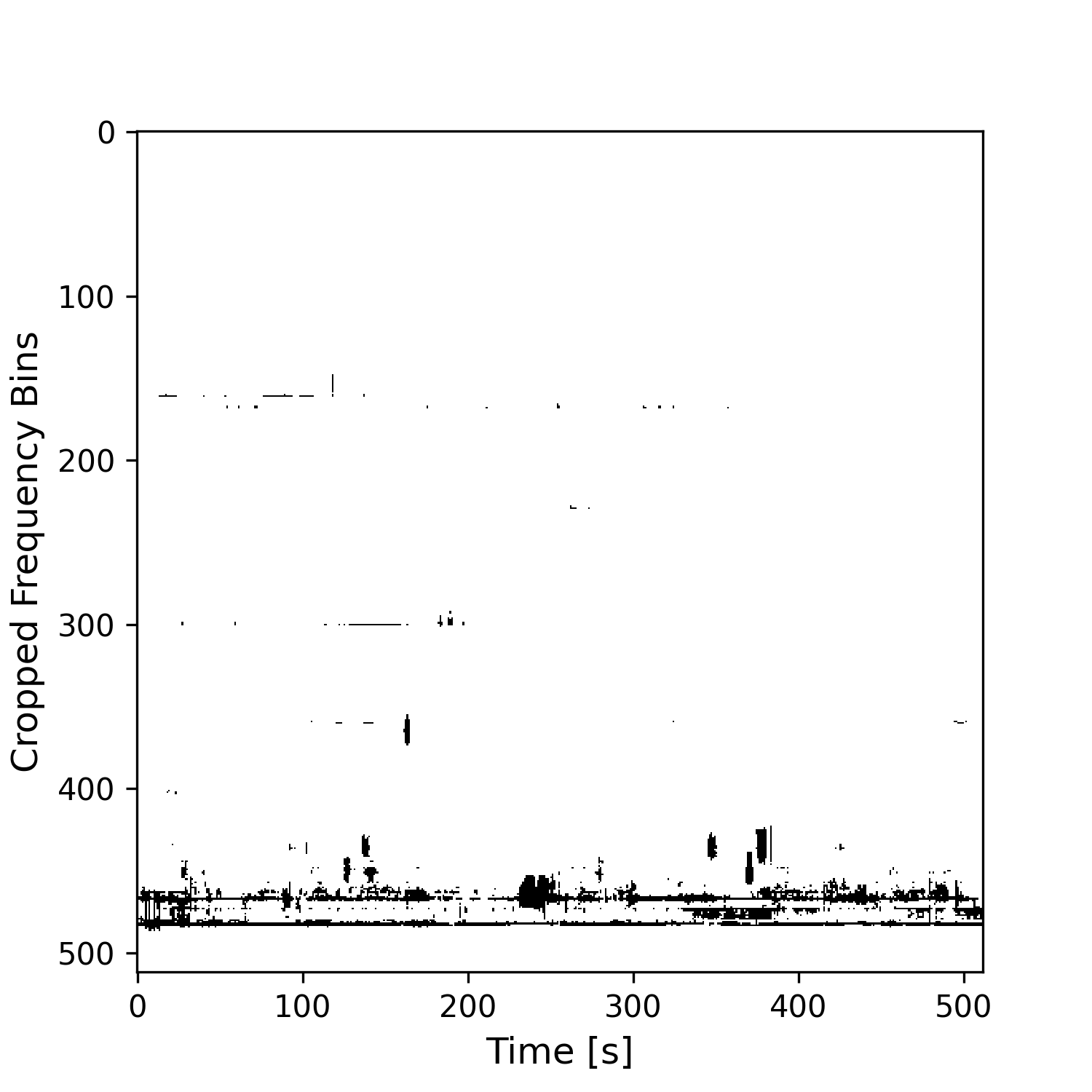}
         \caption{U-Net Prediction}
         \label{fig:95_unet}
     \end{subfigure}
     ~~
      \begin{subfigure}[b]{0.4\linewidth}
         \centering
         \includegraphics[trim={0.1cm 0.1cm 0.1cm 0.1cm},clip,width=0.9\linewidth]{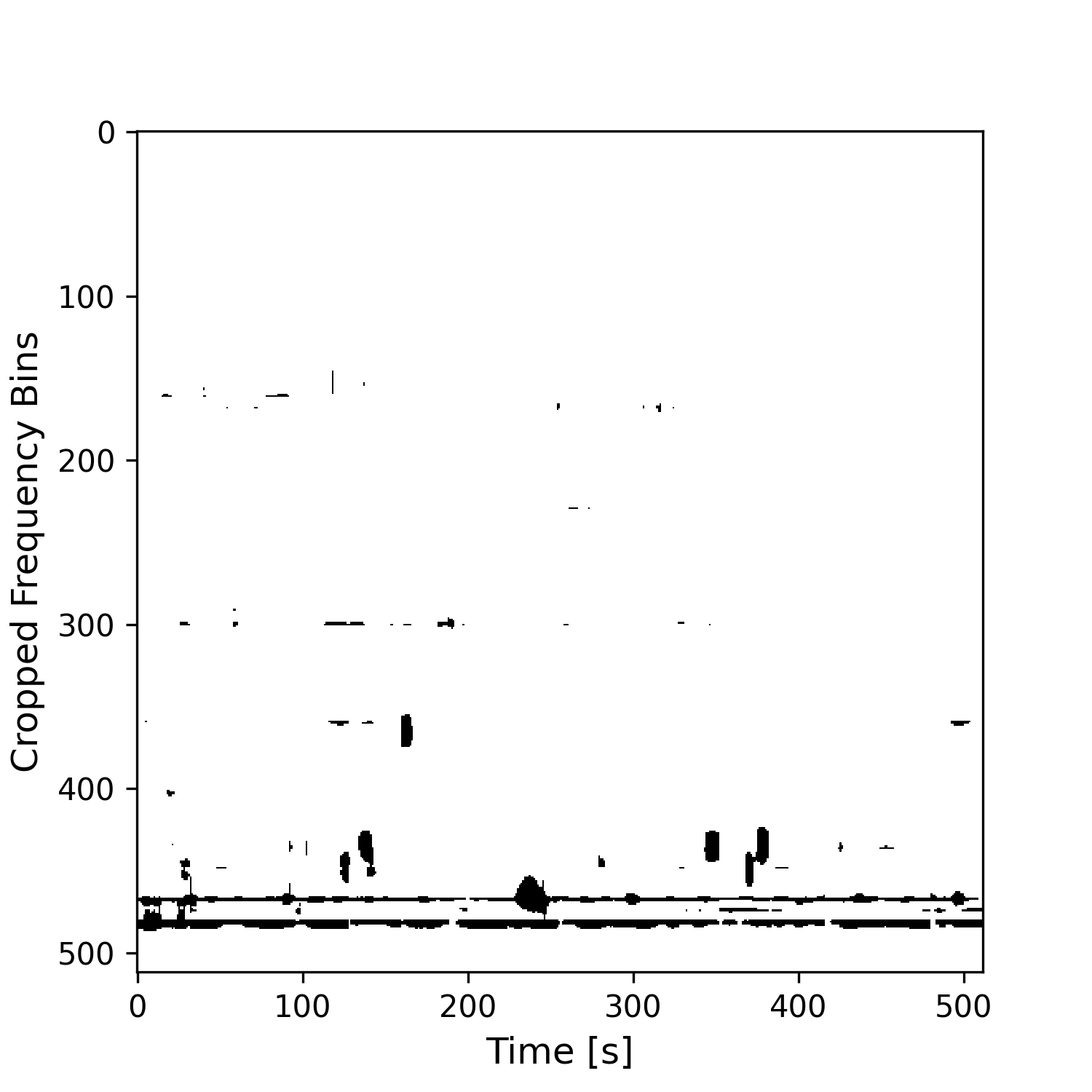}
         \caption{RNet Prediction}
         \label{fig:95_rnet}
     \end{subfigure}
     \\
     \begin{subfigure}[b]{0.4\linewidth}
         \centering
         \includegraphics[trim={0.1cm 0.1cm 0.1cm 0.1cm},width=0.9\linewidth]{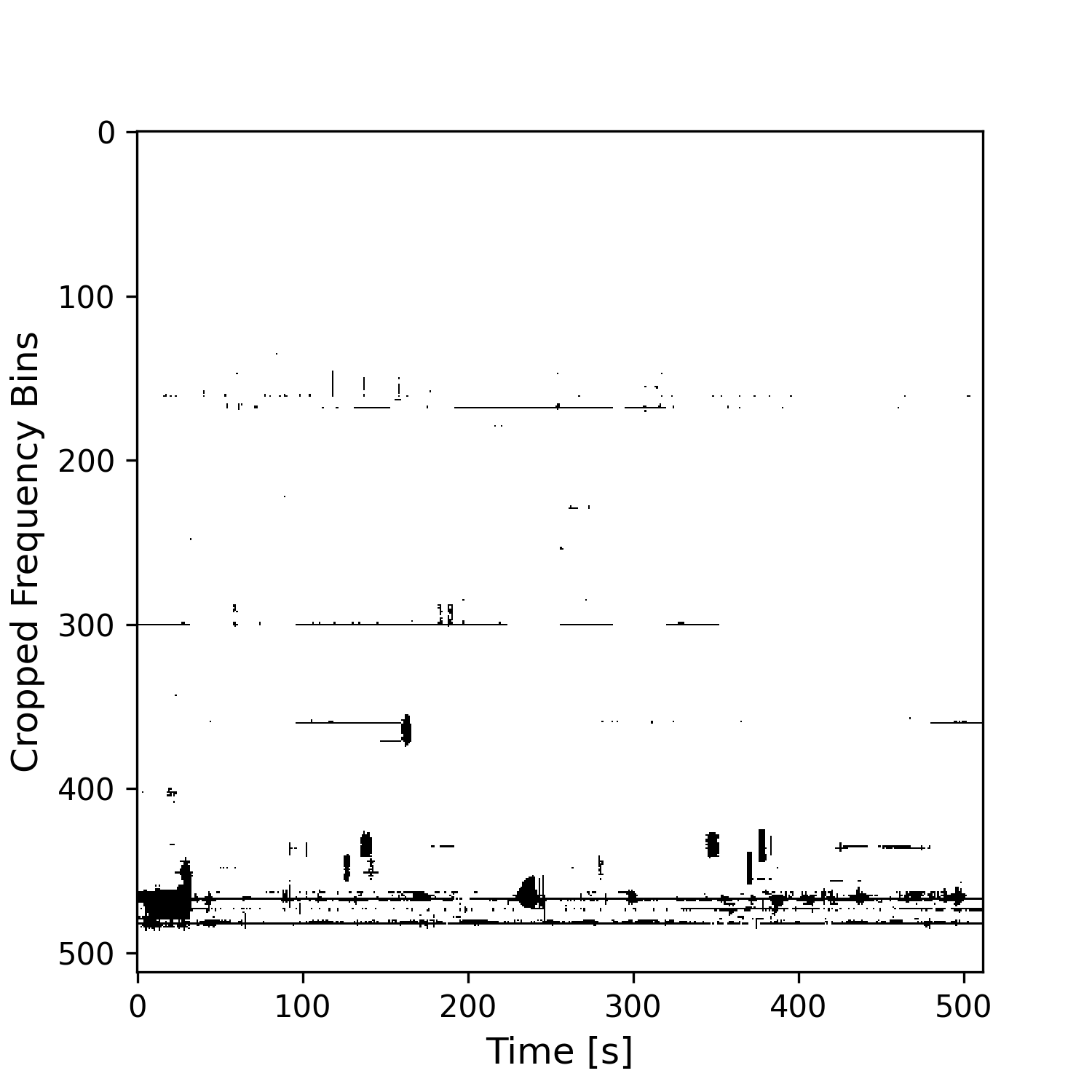}
         \caption{RFI-Net Prediction}
         \label{fig:95_rfinet}
     \end{subfigure}
     ~~
     \begin{subfigure}[b]{0.4\linewidth}
         \centering
         \includegraphics[trim={0.1cm 0.1cm 0.1cm 0.1cm},width=0.9\linewidth]{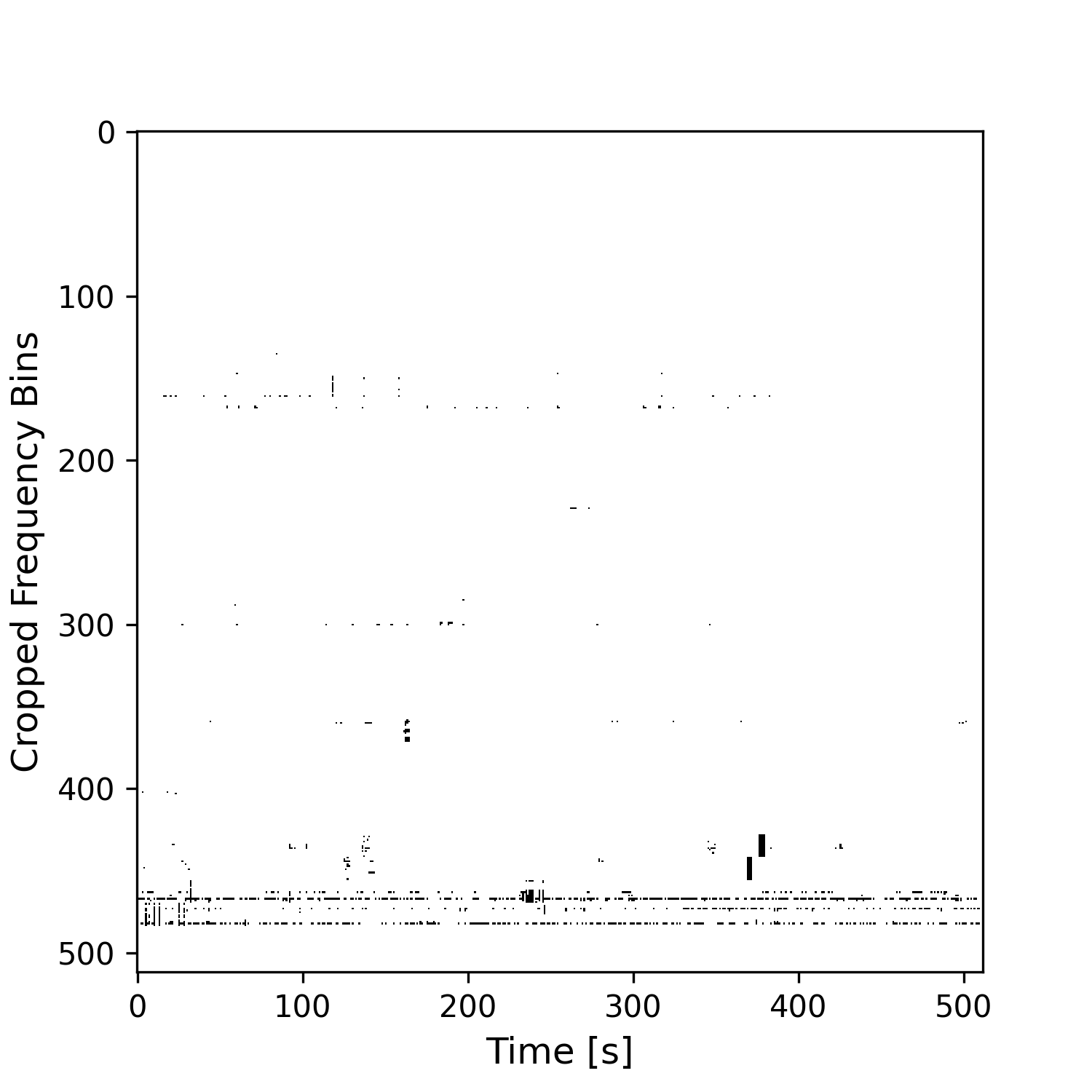}
         \caption{NLN Prediction}
         \label{fig:95_nln}
     \end{subfigure}
        \caption{The 95$^\text{th}$ sample from the testing set, its corresponding mask and the predictions of each model.}
        \label{fig:model_outputs}
\end{figure*}

In Table~\ref{tab:LOFAR_results} we show the performance of NLN relative to the state-of-the-art on the LOFAR dataset. Here we see a similar trend to the HERA-based results; NLN offers superior performance in terms of AUROC and AUPRC. However in terms of F1-score, RFI-Net offers best performance.  The decrease in F1 score is due to the NLN algorithm yielding more false negatives when thresholding all predictions with a single threshold. This is attributable to the large fluctuations in power of the RFI in the LOFAR dataset in combination with the reconstruction-error term of the NLN RFI detector. When the RFI power is low, the reconstruction error will be of a low amplitude and NLN will produce predictions with low power. In effect, when these low power predictions are thresholded using the same threshold as the high power RFI (in this case one which maximises F1-score), the outputs have more false negatives. This is in contrast with the HERA dataset, that has RFI with a consistently higher power level than the astronomical and system-based signals.  A possible solution to this is to predict a threshold on a per spectrogram or patch level, however we consider this out of scope for this research. An illustration of these effects and a comparison between the models is shown in Figure~\ref{fig:model_outputs}.

Furthermore, in the bottom half of Table~\ref{tab:LOFAR_results} we compare NLN with commonly used novelty detection methods. Here we see that NLN significantly outperforms these methods on all metrics. We attribute the performance improvement to the combination of reconstruction error and latent error. 

\begin{figure}
    \centering
    \includegraphics[trim={0.3cm 0.4cm 0.3cm 0.40cm},clip,width=\linewidth]{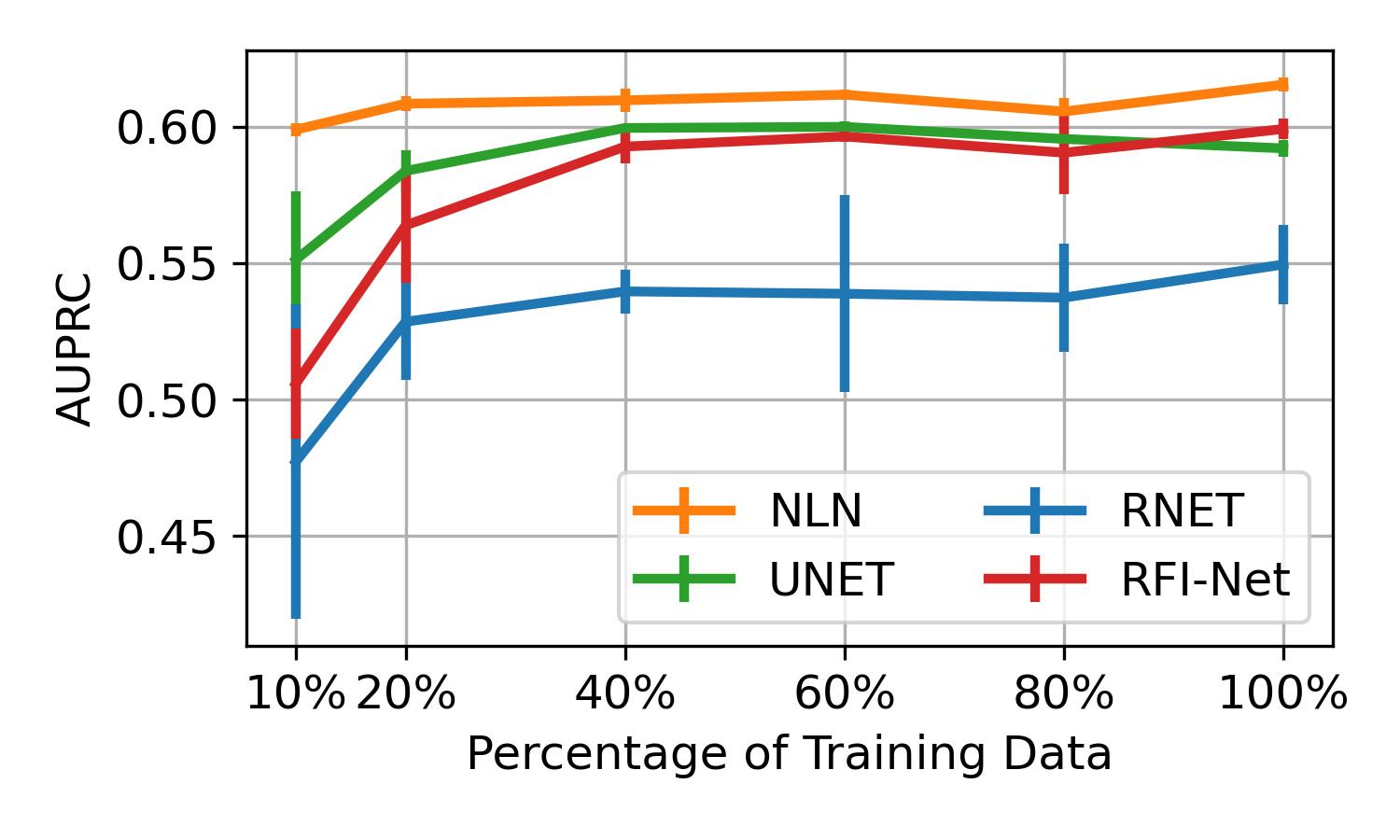}
    \caption{AUPRC performance of each model when training on a percentage of the original LOFAR dataset and evaluating on the original LOFAR test dataset.}
    \label{fig:generalise}
\end{figure}

Additionally, to test the model's reliance on dataset size, we evaluate each model on a percentage of the training data. We show in Figure~\ref{fig:generalise} that NLN is less sensitive to reductions in training data-size, performing almost uniformly with even with large decreases in training data-size. Conversely, the supervised methods' performance scales asymptotically with dataset size, exhibiting significantly higher variance in their performance with smaller dataset sizes. We associate the asymptotic-scaling and increased variance with both the supervised model's larger capacity (due to residual and skip connections) as well as the diversity of the RFI landscape.

\begin{figure*}
     \centering
     \begin{subfigure}[b]{0.3\linewidth}
         \centering
         \includegraphics[trim={0.4cm 0.4cm 0.4cm 0.4cm},clip, width=\linewidth]{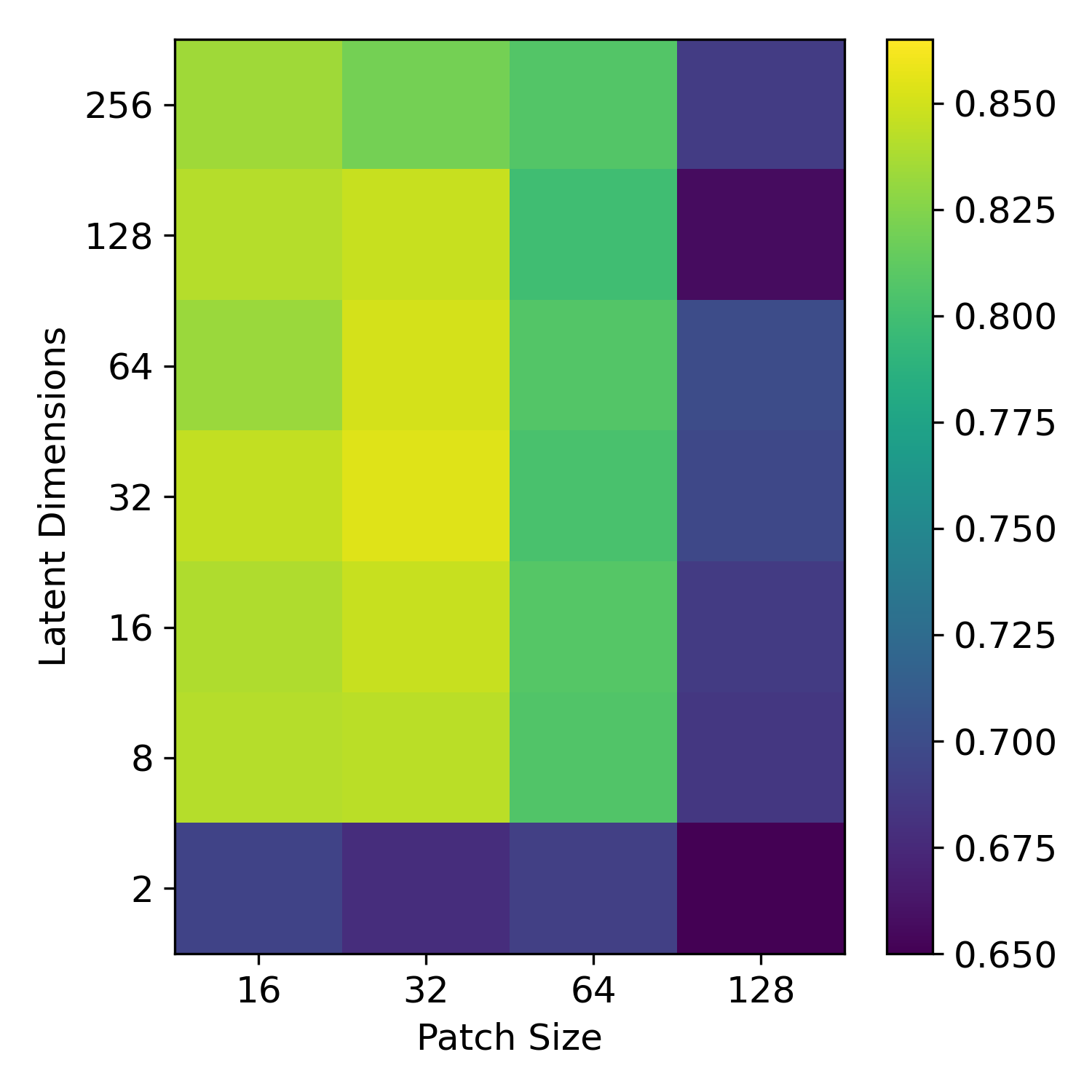}
         \caption{AUROC Sensitivity; \# Neighbours = 16}
         \label{fig:auroc_sen}
     \end{subfigure}
     ~~
     \begin{subfigure}[b]{0.3\linewidth}
         \centering
         \includegraphics[trim={0.4cm 0.4cm 0.4cm 0.4cm},clip,width=\linewidth]{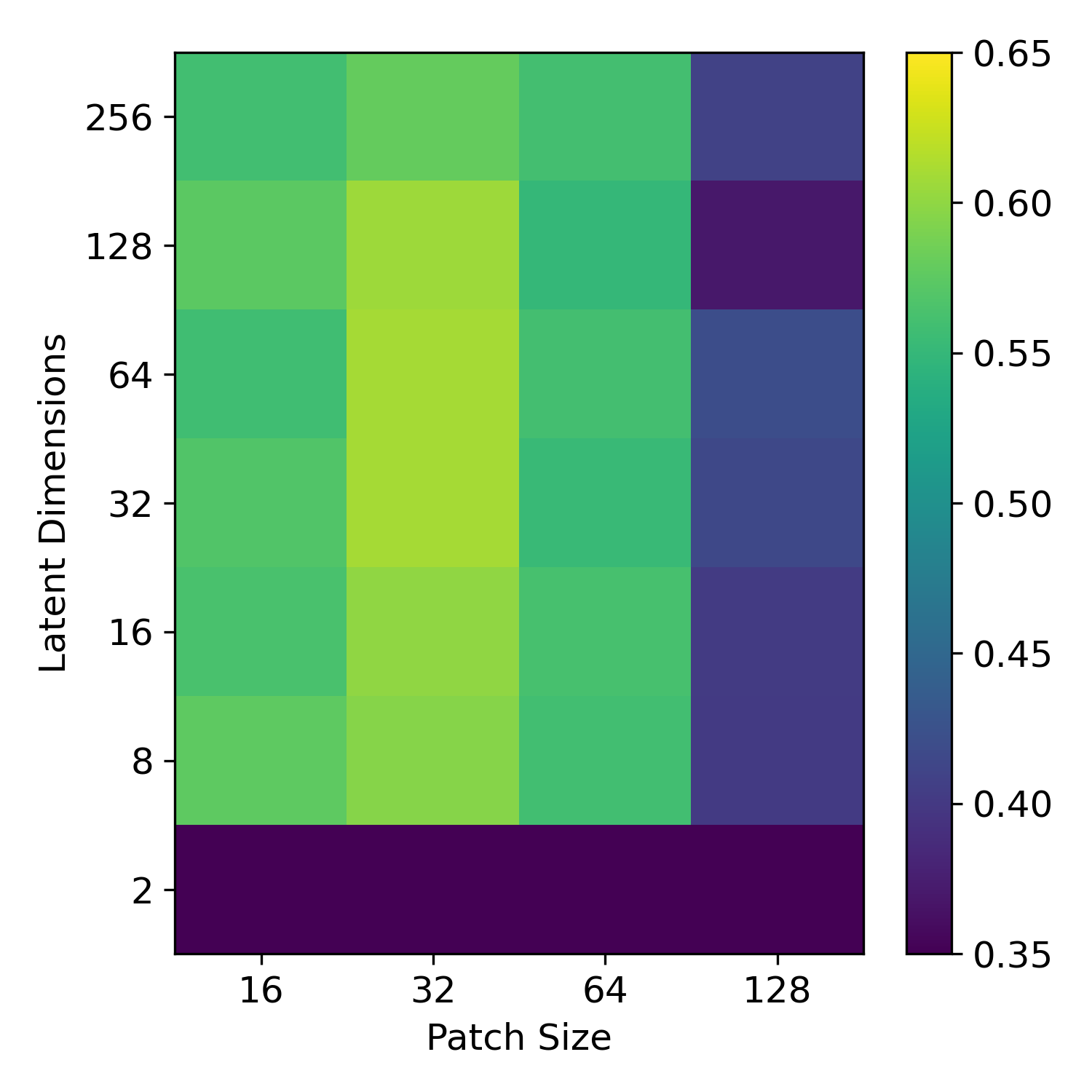}
         \caption{AUPRC Sensitivity; \# Neighbours = 16}
         \label{fig:auprc_sen}
     \end{subfigure}
     ~~
     \begin{subfigure}[b]{0.3\linewidth}
         \centering
         \includegraphics[trim={0.4cm 0.4cm 0.4cm 0.4cm},clip,width=\linewidth]{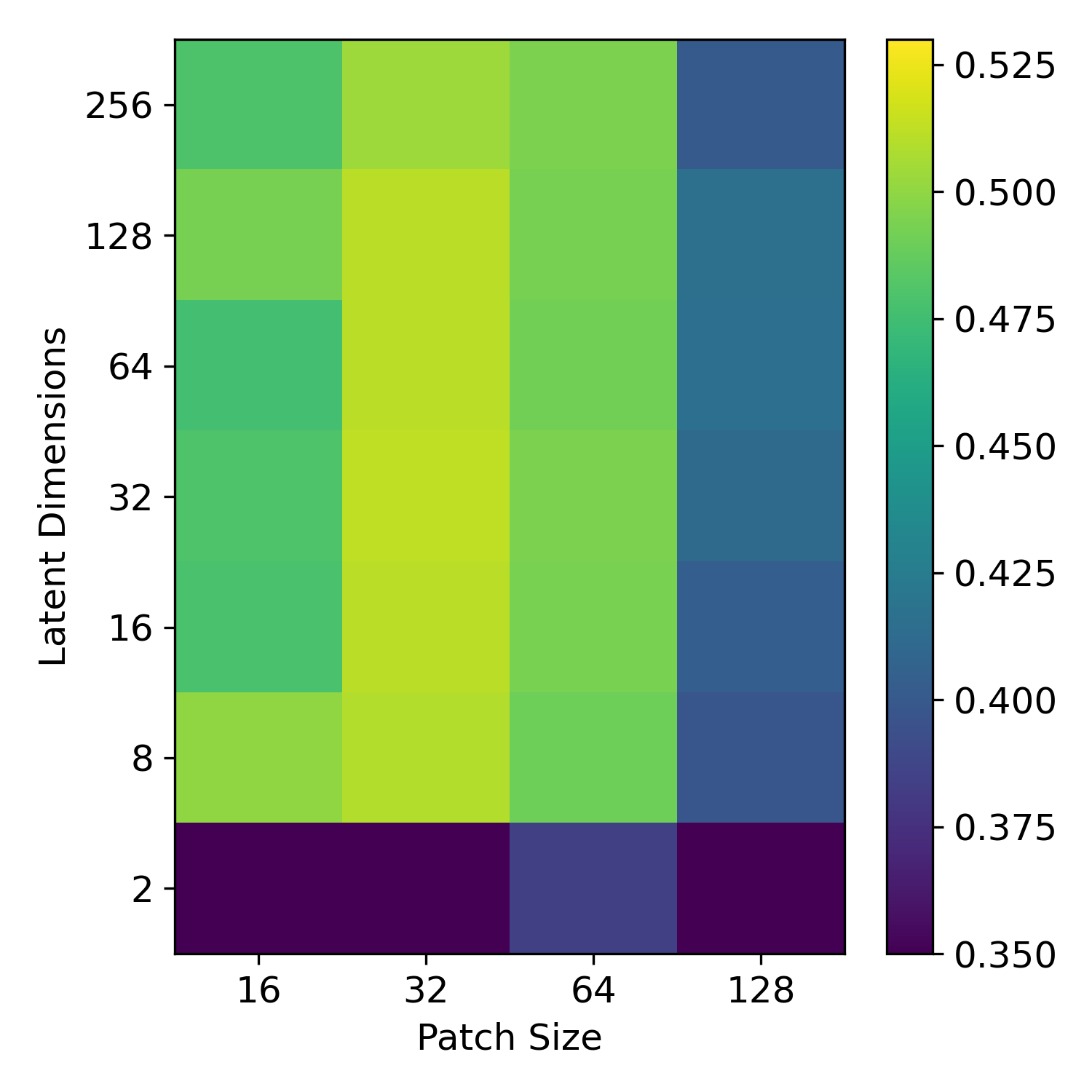}
         \caption{F1 Sensitivity; \# Neighbours = 16}
         \label{fig:f1_sen}
     \end{subfigure}
    \\      
     \begin{subfigure}[b]{0.3\linewidth}
         \centering
         \includegraphics[trim={0.4cm 0.4cm 0.4cm 0.4cm},clip, width=\linewidth]{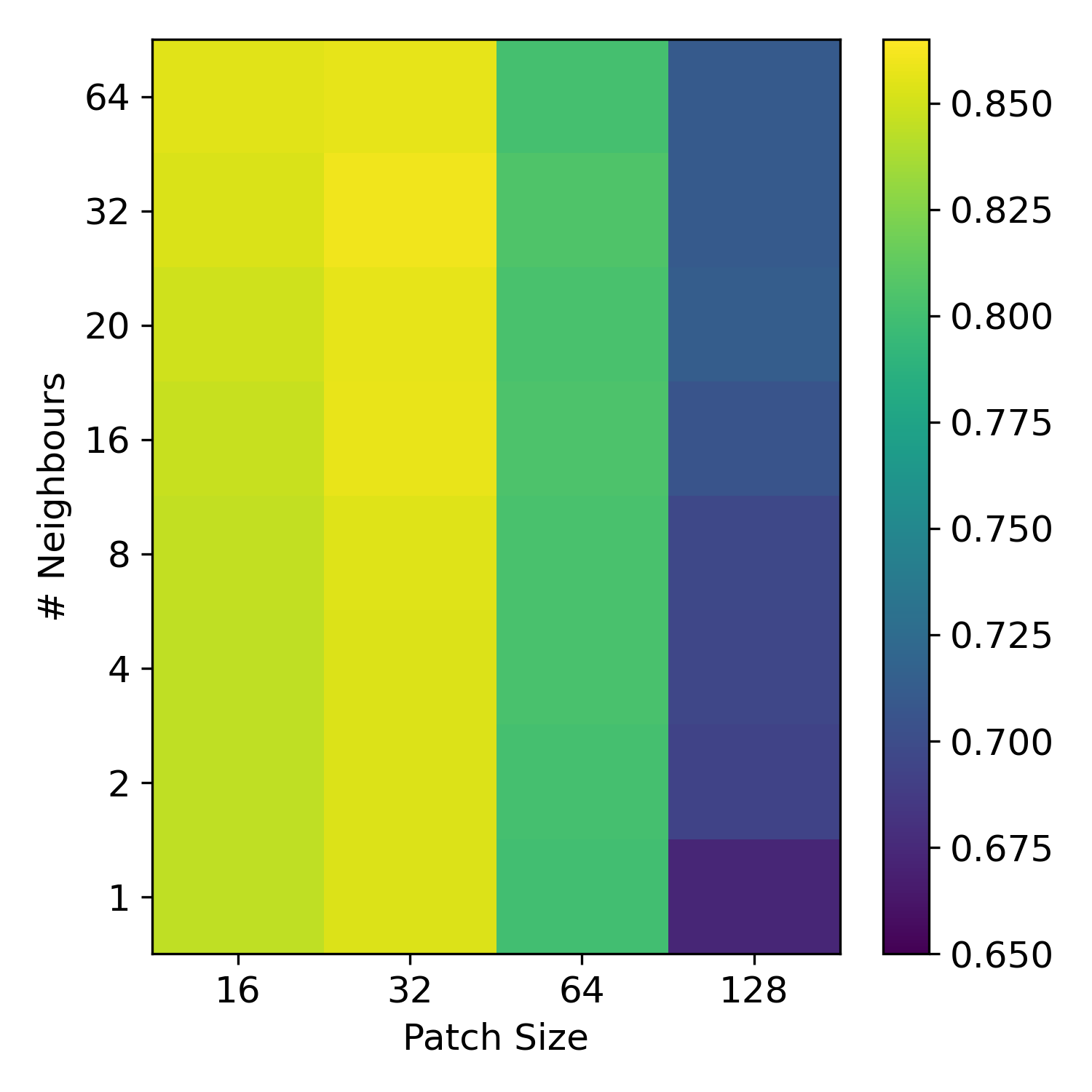}
         \caption{AUROC Sensitivity; Latent dimensions = 32}
         \label{fig:auroc_sen_n}
     \end{subfigure}
     ~~
     \begin{subfigure}[b]{0.3\linewidth}
         \centering
         \includegraphics[trim={0.4cm 0.4cm 0.4cm 0.4cm},clip,width=\linewidth]{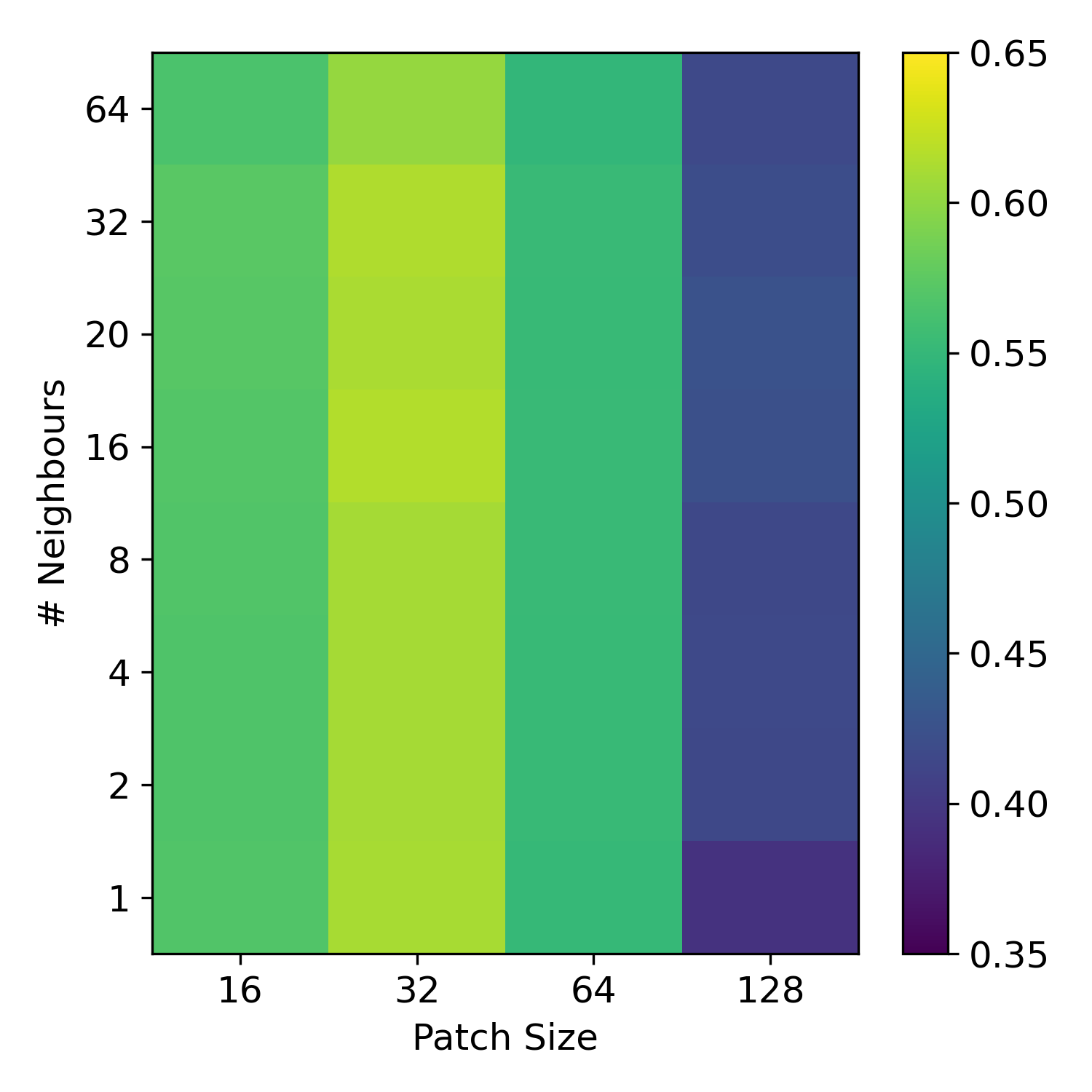}
         \caption{AUPRC Sensitivity; Latent dimensions = 32}
         \label{fig:auprc_sen_n}
     \end{subfigure}
     ~~
     \begin{subfigure}[b]{0.3\linewidth}
         \centering
         \includegraphics[trim={0.4cm 0.4cm 0.4cm 0.4cm},clip,width=\linewidth]{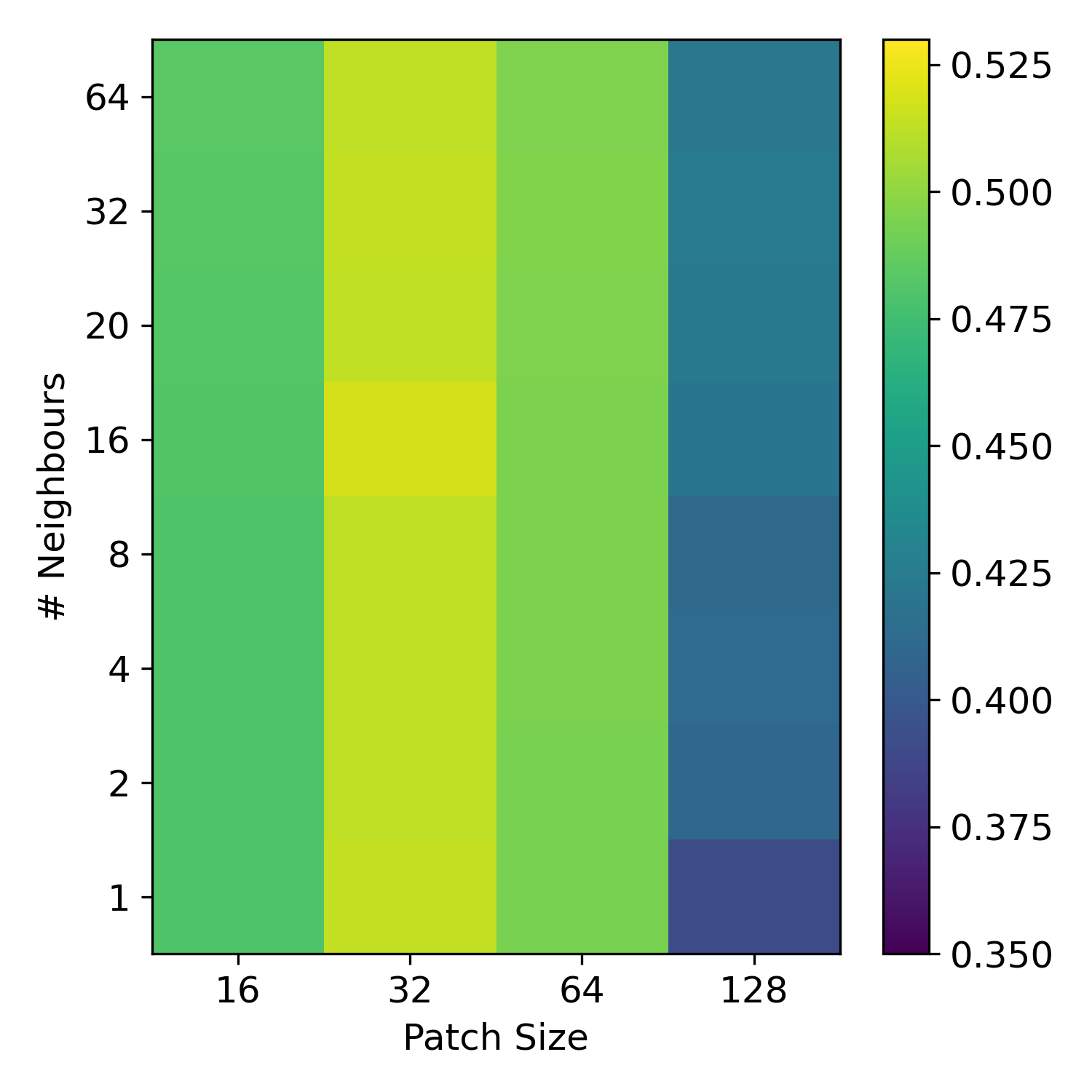}
         \caption{F1 Sensitivity; Latent dimensions = 32}
         \label{fig:f1_sen_n}
     \end{subfigure}

        \caption{Sensitivity of the hyper-parameters on the LOFAR-based performance of NLN when varying latent dimensionality, patch size, and number of neighbours. In order to visualise the four-dimensional space, the number of neighbours is fixed to 16 in the top row, whereas in the bottom we fix the number of latent dimensions to 32. The optimal parameters for the LOFAR dataset are a patch size of $32\times 32$, 32 latent dimensions and 16 neighbours.}
        \label{fig:sensitivty}
\end{figure*}

To determine the sensitivity of the parameters of the NLN-algorithm, we perform a course grid-search of its hyper-parameters. We search across the number of latent dimensions, patch size and number of neighbours as illustrated in Figure~\ref{fig:sensitivty}. In order to better visualise the 4-dimensional space, we plot cross-sections of the high-dimensional landscape. First we fix the number of neighbours to 16 as shown in Figures~\ref{fig:auroc_sen},~\ref{fig:auprc_sen} and~\ref{fig:f1_sen} . It can be seen that the optimal number of latent dimensions is 32, with respect to the average maximum of AUPRC and AUROC. We then set the optimal number of latent dimensions and plot the effect of varying the number of neighbours in Figures ~\ref{fig:auroc_sen_n},~\ref{fig:auprc_sen_n} and~\ref{fig:f1_sen_n}. Through this, we determine the optimal number of neighbours to be 16 with respect to the average maximum AUROC and AUPRC. We conclude that NLN gives optimal performance when the number of latent dimensions are 32, the patch size is $32 \times 32$ and the number of neighbours is 16. 

Finally, as a consequence of our selection RFI-free selection algorithm, we find that NLN requires 66\% less data in comparison to its supervised counterparts. This amount is naturally dataset dependent, however we expect that due to the reduced training data there will be less compute time, and less power consumption while training. 

\section{Conclusions}
\label{sec:Conclusions}

RFI detection is an increasingly important research topic for radio astronomy. State-of-the-art solutions to the RFI problem have been based on supervised machine learning techniques, which fail to address the prohibitive cost of labelling astronomical data. In this work we have documented how inverting the detection problem effectively addresses this issue. We have shown that NLN provides better than state-of-the-art RFI detection without incurring the cost of labelling. 

Furthermore, we have demonstrated that our method better generalises to unseen RFI, whereas current supervised approaches over-fit to weak-label-based RFI masks. As a consequence, we hypothesise that our approach will better generalise to future generations of emitters, whereas existing supervised methods will have to be regularly retrained.  Additionally, we find that due to our training patch selection process we need less data for training, hence decreasing both the training time and energy cost. 

This being said, there is sufficient evidence in several other domains that suggests supervised segmentation algorithms will outperform unsupervised approaches given sufficient high-quality labels.  However, in the current state of the RFI detection domain, where there are still few labelled datasets available and the high cost of obtaining labelled data, we propose the inverted approach as the way forward. 

We plan to further improve the performance of NLN applied to RFI detection through additional training priors. Contrastive self-supervised learning is a  candidate solution thanks to its ability to generate more robust latent representations that can be leveraged in the NLN-algorithm. Furthermore, in order to improve the increased false negative rate of NLN on the LOFAR dataset we suggest more research to be done into automated processing schemes to deal with the high dynamic range of astronomical data when training unsupervised models. This could be additionally improved using a hybrid approach through \texttt{SumThreshold} or trying to directly predict a threshold on a per patch-basis. Finally, we plan to extend this work to more general anomaly detection based problems within radio astronomy.

\section*{Acknowledgements}
This work is part of the "Perspectief" research programme "Efficient Deep Learning" (EDL, \href{https://efficientdeeplearning.nl}{https://efficientdeeplearning.nl}), which is financed by the Dutch Research Council (NWO) domain Applied and Engineering Sciences (TTW). The research makes use of  radio astronomy data from the LOFAR telescope, which is operated by ASTRON (Netherlands Institute for Radio Astronomy), an institute belonging to the Netherlands Foundation for Scientific Research (NWO-I). Special thanks to Prof. dr. ir. Cees T. A. M. de Laat for useful advise and feedback on this research.

\section*{Data Availability}
The data underlying this article are available on Zenodo, at \href{https://dx.doi.org/10.5281/zenodo.6724065}{https://dx.doi.org/10.5281/zenodo.6724065}

\bibliographystyle{mnras}
\bibliography{references.bib} 

\bsp	
\label{lastpage}
\end{document}